\PassOptionsToPackage{unicode}{hyperref}
\PassOptionsToPackage{hyphens}{url}
\PassOptionsToPackage{dvipsnames,svgnames,x11names}{xcolor}
\documentclass[
  10pt,
]{article}
\usepackage{xcolor}
\usepackage[margin=1.25in]{geometry}
\usepackage{amsmath,amssymb}
\usepackage{iftex}
\ifPDFTeX
  \usepackage[T1]{fontenc}
  \usepackage[utf8]{inputenc}
  \usepackage{textcomp} 
\else 
  \usepackage{unicode-math} 
  \defaultfontfeatures{Scale=MatchLowercase}
  \defaultfontfeatures[\rmfamily]{Ligatures=TeX,Scale=1}
\fi
\usepackage{lmodern}
\ifPDFTeX\else
\fi
\IfFileExists{upquote.sty}{\usepackage{upquote}}{}
\IfFileExists{microtype.sty}{
  \usepackage[]{microtype}
  \UseMicrotypeSet[protrusion]{basicmath} 
}{}
\makeatletter
\@ifundefined{KOMAClassName}{
  \IfFileExists{parskip.sty}{%
    \usepackage{parskip}
  }{
    \setlength{\parindent}{0pt}
    \setlength{\parskip}{6pt plus 2pt minus 1pt}}
}{
  \KOMAoptions{parskip=half}}
\makeatother
\usepackage{longtable,booktabs,array}
\usepackage{calc} 
\usepackage{etoolbox}
\makeatletter
\patchcmd\longtable{\par}{\if@noskipsec\mbox{}\fi\par}{}{}
\makeatother
\IfFileExists{footnotehyper.sty}{\usepackage{footnotehyper}}{\usepackage{footnote}}
\makesavenoteenv{longtable}
\usepackage{graphicx}
\makeatletter
\newsavebox\pandoc@box
\newcommand*\pandocbounded[1]{
  \sbox\pandoc@box{#1}%
  \Gscale@div\@tempa{\textheight}{\dimexpr\ht\pandoc@box+\dp\pandoc@box\relax}%
  \Gscale@div\@tempb{\linewidth}{\wd\pandoc@box}%
  \ifdim\@tempb\p@<\@tempa\p@\let\@tempa\@tempb\fi
  \ifdim\@tempa\p@<\p@\scalebox{\@tempa}{\usebox\pandoc@box}%
  \else\usebox{\pandoc@box}%
  \fi%
}
\def\fps@figure{htbp}
\makeatother
\NewDocumentCommand\citeproctext{}{}

\makeatletter
 \let\@cite@ofmt\@firstofone
 \def\@biblabel#1{}
 \def\@cite#1#2{{#1\if@tempswa , #2\fi}}
\makeatother
\newlength{\cslhangindent}
\newlength{\csllabelwidth}
\newenvironment{CSLReferences}[2] 
 {\begin{list}{}{%
  \setlength{\itemindent}{0pt}
  \setlength{\leftmargin}{0pt}
  \setlength{\parsep}{0pt}
  \ifodd #1
   \setlength{\leftmargin}{\cslhangindent}
   \setlength{\itemindent}{-1\cslhangindent}
  \fi
  \setlength{\itemsep}{#2\baselineskip}}}
 {\end{list}}
\usepackage{calc}

\providecommand{\tightlist}{%
  \setlength{\itemsep}{0pt}\setlength{\parskip}{0pt}}
\usepackage{booktabs}
\usepackage{caption}
\usepackage{bookmark}
\IfFileExists{xurl.sty}{\usepackage{xurl}}{} 
\hypersetup{
  pdftitle={An ICLR Acceptance Says Less Than It Used To},
  pdfauthor={Mikhail L. Arbuzov; Sisong Bei; Lee Mosbacker},
  colorlinks=true,
  linkcolor={Maroon},
  filecolor={Maroon},
  citecolor={Blue},
  urlcolor={Blue},
  pdfcreator={LaTeX via pandoc}}

\title{An ICLR Acceptance Says Less Than It Used To}
\author{Mikhail L. Arbuzov \and Sisong Bei \and Lee Mosbacker}
\date{}

\begin{document}
\maketitle
\begin{abstract}
Hiring committees, funders and promotion panels treat an ICLR acceptance
as evidence that a paper is good, and researchers pay for that evidence
with their time: by early November, we estimate, the first round of
reviews for ICLR 2027 had taken 188 working years. We show that an
acceptance now tells them less about a paper than it did five years ago.
From 2023 to 2027 the number of reviewed papers grew eightfold, and
ICLR's public first-round ratings became less able to tell those papers
apart. Through 2025, reviewers of the same paper disagreed about as much
as before, but fewer papers received clearly high ratings and more of
them clustered near the acceptance cutoff, where an ordinary difference
of opinion decides which paper gets a place. In simulations fitted to
each year's ratings, a paper that would rank in the top tenth over many
independent reviews was rejected one time in eight in 2023 and one time
in four in 2027. Rebuttal and discussion then bring the scores closer
together, but agreement is not accuracy: when we simulated those score
changes without adding new information about the papers, reviewers
agreed more and the accepted list got worse. Nor are more reviews a
remedy, since giving each 2027 paper as many reviews as in 2026 would
take about 50 more working years and improve selection only slightly. As
submissions keep growing, conferences cannot sustain this process;
before adopting a new one, they should estimate which papers it would
select and how much reviewing it would cost.
\end{abstract}

\section{The bill, and what rides on
it}\label{the-bill-and-what-rides-on-it}

Ask around at ICLR and you will hear that it has become a lottery. Most
people who submit there have the story: the good paper that was
rejected, the weaker one that got in. Whether that is true, and how it
came about, matters because of what the review costs.

By early November 2026, the 40,458 submissions to ICLR 2027 with public
ratings had drawn 125,048 reviews ({shiqiangw} 2026). At three hours a
review, our working figure, that is about 375,000 hours of qualified
researchers' time: some 188 working years at 2,000 hours a year, for the
first round of one conference in one year. It comes on top of the months
each of those papers took to write.

In return, some authors receive an acceptance, and hiring committees,
funders and promotion panels read it as a statement about the work. If
that statement is getting weaker, the field is spending more and more of
its best people's time to buy less and less. We wanted to know whether
it is, and what is going on. We looked at five years of public
first-round ratings, 2023 to 2027, and modelled where papers would land
if those scores decided acceptance.

We think it is getting weaker. The first reading tells less about the
paper each year. Better papers still do better, but when our model
selects about three reviewed papers in ten, as ICLR did in 2024 and
2025, the line in 2027 runs through a crowd that the first reading can
scarcely tell apart. Run the 2027 conference in our model a second time
with fresh reviewers, and less than half the accepted list comes back.
Low first-round scores also predict which papers leave. By 2025 a
quarter of reviewed papers were recorded as withdrawn. On the papers
decided in 2024 and 2025, rebuttal and discussion brought the scores
closer together, mostly by raising low ratings. When we simulated those
changes using the first-round scores alone, reviewers agreed more and
the accepted list got worse. What the reviewers learned from the actual
rebuttals is something we cannot read from their scores.

In 2027, in the ratings we have, the flood of submissions was met with a
coarser scale and fewer reviews per paper, and the model says these did
most of the year's damage. We read that as what patching a strained
process under load looks like. Patching will not bring back what an
acceptance used to say about a paper. We have to treat peer review as a
decision process and work out what a design costs and where it places
papers before adopting it.

\section{Where papers land}\label{where-papers-land}

In a conference of ICLR 2027's size and noise, good papers fall below
the line and weak ones rise above it. Selection is real. At the line,
though, the first reading tells too little.

Nobody observes a paper's true quality, so nobody can count the good
papers a conference rejected. We use a narrower measure. A paper's
quality is whatever its reviewers share, and a paper is good, or belongs
on the accepted list, when it is in the top 32\% by quality.

What the public ratings do show, for every year, is how far the reviews
of one paper differ from each other against how far papers differ from
one another. We call that ratio the review noise. The model behind this
paper reads it from the counts of ratings alone; it is the model of our
earlier paper (Arbuzov et al. 2026), set out in Appendix A. For ICLR
2027 the noise is 4.21, which means that two underlying readings of one
paper correlate 0.19, where 1 would mean the same review and 0 an
unrelated one. We then simulate a conference of 40,458 papers whose
reviews behave that way, with ICLR 2027's numbers of reviews and its
four-level scale. It accepts the top 32\% by mean rating, the share ICLR
accepted in 2024 and 2025, and we read off who gets in.

\begin{figure}
\centering
\includegraphics[width=0.84\linewidth,height=\textheight,keepaspectratio,alt={One simulated ICLR 2027. Each dot is a paper, placed by its quality (left to right) and the mean rating it received (up). Blue papers were accepted and grey ones rejected; the horizontal line is the accept line, and the dashed line marks the top 32\% by quality. The right panel counts the papers at each quality.}]{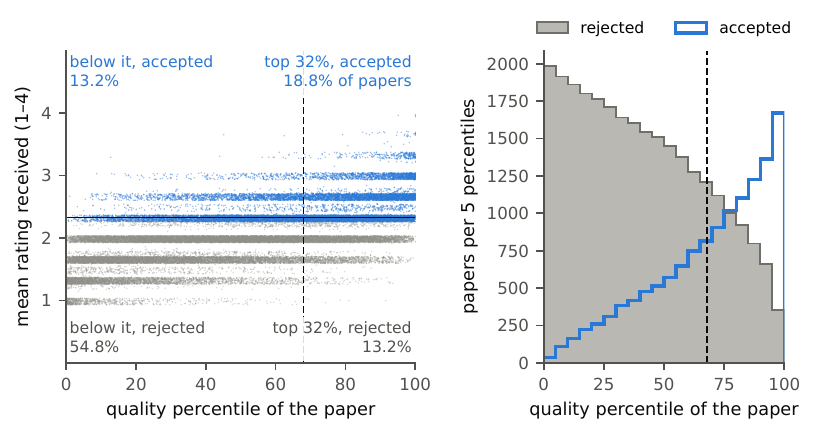}
\caption{One simulated ICLR 2027. Each dot is a paper, placed by its
quality (left to right) and the mean rating it received (up). Blue
papers were accepted and grey ones rejected; the horizontal line is the
accept line, and the dashed line marks the top 32\% by quality. The
right panel counts the papers at each quality.}
\end{figure}

Follow any band of ratings across Figure 1 and papers of every quality
are in it. Of the papers accepted, 41\% do not belong on the list, and
22\% come from the bottom half of the pool. Run the same conference
again with fresh reviewers, and only 47\% of the accepted list returns;
accepting at random would return 32\%.

Seen from a paper's side, the same picture is a risk. Of the papers that
do belong in the top 32\%, two in five are turned away. A paper in the
top tenth by quality is rejected one time in four, and a paper right at
the line is accepted four times in ten.

It is not a lottery, though. The median accepted paper sits at the 75th
percentile of quality and the median rejected paper at the 38th: better
papers do better, and the weakest are usually rejected. What fails is
the line. A paper has to be in the top sixth by quality before its
expected mean rating clears it, and the mean rating it actually gets
scatters around that expectation with the luck of its reviewers. For
about six papers in ten, those from the 38th percentile of quality to
the 99th, the line is closer to a paper's expected mean rating than the
typical scatter around it (Appendix C). Across that stretch a rating is
a weak signal of which side of the line a paper belongs on.

\section{Five years of first-round
ratings}\label{five-years-of-first-round-ratings}

The estimated first-round signal fell in each year from 2023 to 2027,
while the pool of reviewed papers grew from 4,833 to 40,458. In the
simulation, a top-tenth paper's risk of rejection rose, and so did the
share of accepted papers that did not belong.

We fitted the same model to the first-round ratings of each year,
including rated withdrawals wherever the source records them (Appendix
B). In the model, two underlying readings of the same paper correlated
0.27 in 2023, then 0.23, 0.22, 0.20 and 0.19. In terms of noise, that is
a rise from 2.73 to 4.21 (Figure 2). The estimate is higher each year;
most neighbouring years' margins of error overlap, but the rise over the
five years lies far outside them. The first three years shared one
rating scale, so the rise from 2.73 to 3.58 is not an effect of the
scale. Two of the five years are snapshots: the 2026 and 2027 ratings
come from a rollback and an early-discussion snapshot, respectively
(Appendix B).

\begin{figure}
\centering
\includegraphics[width=0.64\linewidth,height=\textheight,keepaspectratio,alt={First-round review noise by year, with 95\% intervals from 300 resamples (blue), beside the noise of final ratings, 2018 to 2025 (grey). At noise 2.73 two underlying readings of a paper correlate 0.27, at 4.21 they correlate 0.19. The public ratings for 2026 and 2027 are not final ratings (Appendix B).}]{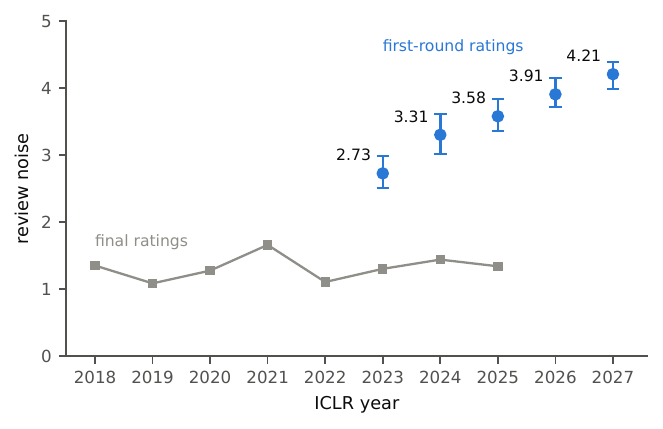}
\caption{First-round review noise by year, with 95\% intervals from 300
resamples (blue), beside the noise of final ratings, 2018 to 2025
(grey). At noise 2.73 two underlying readings of a paper correlate 0.27,
at 4.21 they correlate 0.19. The public ratings for 2026 and 2027 are
not final ratings (Appendix B).}
\end{figure}

The final ratings show a different pattern. Their noise stays between
1.1 and 1.7 in every year from 2018 to 2025, with no detectable linear
trend. Final scores alone would miss the first-round rise over the years
the series overlap.

Table 1 runs the simulated conference at each year's noise, with the
year's own reviews per paper and rating scale, accepting the top 32\%
every time.

\begin{longtable}[]{@{}
  >{\raggedright\arraybackslash}p{(\linewidth - 10\tabcolsep) * \real{0.1304}}
  >{\raggedleft\arraybackslash}p{(\linewidth - 10\tabcolsep) * \real{0.1739}}
  >{\raggedleft\arraybackslash}p{(\linewidth - 10\tabcolsep) * \real{0.1739}}
  >{\raggedleft\arraybackslash}p{(\linewidth - 10\tabcolsep) * \real{0.1739}}
  >{\raggedleft\arraybackslash}p{(\linewidth - 10\tabcolsep) * \real{0.1739}}
  >{\raggedleft\arraybackslash}p{(\linewidth - 10\tabcolsep) * \real{0.1739}}@{}}
\caption{The same conference at each year's noise. ``Belongs'' means in
the top 32\% by quality; ``a rerun returns'' is the share of the
accepted list that fresh reviewers would accept again. These are model
figures, not counts.}\tabularnewline
\toprule\noalign{}
\begin{minipage}[b]{\linewidth}\raggedright
Year
\end{minipage} & \begin{minipage}[b]{\linewidth}\raggedleft
Reviewed papers
\end{minipage} & \begin{minipage}[b]{\linewidth}\raggedleft
Latent reading correlation
\end{minipage} & \begin{minipage}[b]{\linewidth}\raggedleft
Accepted list that belongs
\end{minipage} & \begin{minipage}[b]{\linewidth}\raggedleft
A rerun returns
\end{minipage} & \begin{minipage}[b]{\linewidth}\raggedleft
Top-tenth paper rejected
\end{minipage} \\
\midrule\noalign{}
\endfirsthead
\toprule\noalign{}
\begin{minipage}[b]{\linewidth}\raggedright
Year
\end{minipage} & \begin{minipage}[b]{\linewidth}\raggedleft
Reviewed papers
\end{minipage} & \begin{minipage}[b]{\linewidth}\raggedleft
Latent reading correlation
\end{minipage} & \begin{minipage}[b]{\linewidth}\raggedleft
Accepted list that belongs
\end{minipage} & \begin{minipage}[b]{\linewidth}\raggedleft
A rerun returns
\end{minipage} & \begin{minipage}[b]{\linewidth}\raggedleft
Top-tenth paper rejected
\end{minipage} \\
\midrule\noalign{}
\endhead
\bottomrule\noalign{}
\endlastfoot
2023 & 4,833 & 0.27 & 67.5\% & 56.4\% & 13\% \\
2024 & 6,954 & 0.23 & 65.7\% & 54.3\% & 15\% \\
2025 & 11,423 & 0.22 & 64.8\% & 53.6\% & 16\% \\
2026 & 18,791 & 0.20 & 63.9\% & 52.4\% & 17\% \\
2027 & 40,458 & 0.19 & 58.5\% & 47.0\% & 25\% \\
\end{longtable}

A paper in the top tenth was rejected one time in eight in 2023 and one
time in four in 2027, and the share of the accepted list that belongs
fell from two in three to under three in five. The list loosened
gradually through 2026 (3.6 points) and then lost 5.4 points in 2027.

Noise is a ratio, and it can rise in two ways: reviews of a paper can
scatter more, or papers can draw closer together. The next section shows
which happened in 2023 to 2025.

\section{Why: the clear accepts thinned
out}\label{why-the-clear-accepts-thinned-out}

The first suspicion is careless reviewers, overloaded by the growth of
the pool. The ratings say otherwise. Through 2025, reviewers of one
paper disagreed about as much as they did in 2023. What changed is the
top of the ratings: the clear accepts thinned out, the papers bunched on
the line, and the distance between papers shrank. The agreement
advantage of reviewers who rated themselves confident was smallest in
2025.

Between ICLR 2023 and 2025 the rating scale and the review form did not
change. On that scale (1, 3, 5, 6, 8 and 10) a 5 means marginally below
the acceptance threshold, a 6 marginally above it, and an 8 a clear
accept. Across those years the variance among the ratings of one paper
slipped from 2.27 to 2.05, the confidence reviewers gave themselves held
at 3.6 on a scale of 5, and the share of low ratings (1 or 3) stayed
near a third. The reviews of a single paper were no more scattered than
before.

What changed is at the top, and Figure 3 shows it. Ratings of 8 fell
from 12.0\% of all ratings to 7.8\%, and ratings of 5 rose from 32.3\%
to 35.6\%. By 2025 six ratings in ten were a 5 or a 6, the two marginal
ratings on either side of the threshold. The share of papers averaging 6
or more fell from 21\% to 13\%, although the pool grew 2.4 times; the
papers that stood out grew by half. Top ratings were less common in 2025
than in 2023 at every level of stated confidence, so it is not one kind
of reviewer.

\begin{figure}
\centering
\includegraphics[width=0.52\linewidth,height=\textheight,keepaspectratio,alt={The share of each rating among first-round ratings at ICLR 2023, 2024 and 2025, which share one scale. A 5 is marginally below the acceptance threshold, a 6 marginally above it, and an 8 a clear accept.}]{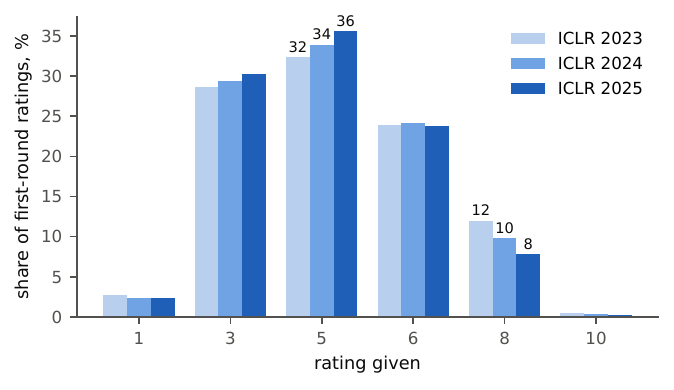}
\caption{The share of each rating among first-round ratings at ICLR
2023, 2024 and 2025, which share one scale. A 5 is marginally below the
acceptance threshold, a 6 marginally above it, and an 8 a clear accept.}
\end{figure}

That is how the noise rises with no one scattering more. Noise sets the
spread among a paper's reviews against the spread between papers, and
the second shrank: the estimated between-paper variance fell by a third,
from 0.73 to 0.49 (Appendix D). The same disagreement among reviewers
now swallows more of the distance between papers, and a review tells
less about where a paper stands among the others.

The confident reviewers show it most. In 2023 two reviewers who both
said they were confident (4 or 5 out of 5) agreed on a paper clearly
better than two who were not: their noise was 2.7, against 4.0. By 2025
it was 3.9 against 5.1 (Appendix D). The within-paper variance of their
ratings fell by a tenth (2.34 to 2.12) while the estimated variance
between the papers they rated fell by more than a third (0.86 to 0.54).
Their agreement fell as everyone's did, because the papers drew
together.

We cannot tell from the ratings why the top thinned. The new submissions
may be mostly middling papers, so that standout work became a shrinking
share; or reviewers may have grown reluctant to give an 8. The low end
of the ratings did not move, which argues a little against reviewers
simply growing harsher. For sorting papers the consequence is the same
either way: the papers are now bunched around the line.

\section{Rebuttal and discussion align the
scores}\label{rebuttal-and-discussion-align-the-scores}

The natural objection is that the first round is only a first look. Most
ratings stay put through rebuttal and discussion, but reviewers who
started below the others are especially likely to raise theirs. The list
has a fixed number of places, so each paper brought onto it pushes
another off. When we replayed those changes using first-round scores
alone, reviewers agreed more and roughly a fifth of the score-ranked
list changed. We could check both effects against the real scores, but
only the replay could tell us whether the papers that came in were
better than the ones they displaced.

Among papers that reached a decision, most ratings never change, and
fewer of them go unchanged each year: 81\% in 2023, 69\% in 2025. Nearly
every change is upward: in 2025, 30\% of ratings were raised and 1.5\%
lowered (Appendix E); Kargaran et al. (2025) report the same one-way
pattern for 2024 and 2025. The ratings that rise are the low-standing
ones (Figure 4). In 2025 half of the ratings that stood more than 2.5
points under the paper's other reviews were raised; of those standing at
least 2.5 points above them, one in twenty was lowered, and almost none
raised.

\begin{figure}
\centering
\includegraphics[width=0.58\linewidth,height=\textheight,keepaspectratio,alt={For ICLR 2024 and 2025, the share of ratings raised (up) and lowered (down) after the first round, by where the rating stood among the paper's other reviews, in rating points. Papers not withdrawn, with three or more reviews.}]{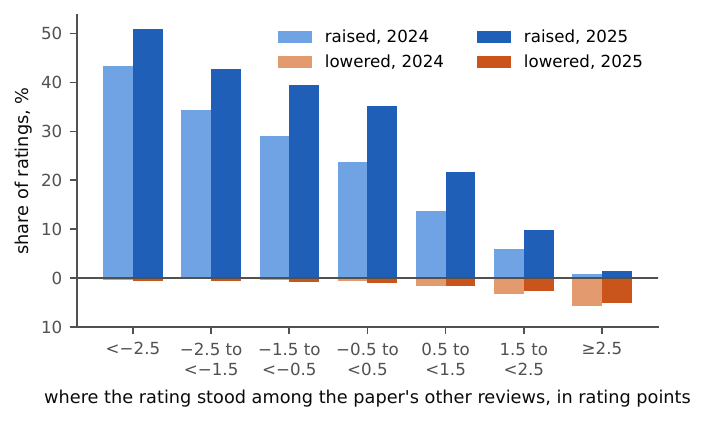}
\caption{For ICLR 2024 and 2025, the share of ratings raised (up) and
lowered (down) after the first round, by where the rating stood among
the paper's other reviews, in rating points. Papers not withdrawn, with
three or more reviews.}
\end{figure}

The raising is clearest for a paper with a lone low review, at least two
points below each of the others, which themselves span at most two
points. About three in ten papers that stayed in had one. It was raised
in half of them, and a paper whose lone low review rose was accepted
53\% of the time, against 19\% when it did not. In the other direction
there is almost nothing. In 2023 we paired ratings by their position in
each paper's review list (Appendix B). On papers that ended up accepted,
ratings went up about seventy times for every one that went down. On
rejected papers the ratio was under four to one.

Most of the accepted list still comes from the first reading. Of the
papers leading by first-round mean, as many as were accepted, 79\%, 75\%
and 75\% were accepted in 2023, 2024 and 2025. The rest of the list, a
fifth to a quarter, comes from below the first-round cutoff.

The agreement that results is large. On the recorded ratings of papers
that were not withdrawn, two reviews of one paper correlate 0.33 after
the process, against 0.15 before in 2025 (0.17 in 2024): about twice as
much. These are recorded ratings, not the model's figures of Table 1.
Where does the agreement come from? We took the recorded first-round
ratings, replaced some with their actual final values, and measured
agreement again. Letting only the ratings on papers that ended up
accepted take their final values reproduces nearly all of the gain;
letting only those on rejected papers change leaves agreement near its
first-round level (Appendix E). The papers move apart, the variance
between them doubling, far more than the reviewers of a paper close
ranks. The gain in agreement is concentrated on eventual acceptances.
The records do not say whether a score change influenced the decision or
followed it.

What would alignment alone do? We fitted the model to the decided papers
with paired reviews in 2024 and 2025. For each simulated review, we
independently drew a final rating from real reviews with the same
initial rating and a similar distance from the other reviewers' mean. In
2024, agreement rose from 0.17 to 0.27 in the replay and from 0.17 to
0.33 in the data. In 2025 it rose from 0.15 to 0.26 and from 0.15 to
0.33. At a fixed 32\% cutoff, the replay replaced 19\% and 21\% of the
first-round lists, against 23\% and 26\% in the observed score rankings
(Appendix E).

The replay made selection worse. In 2024 the share of selected papers
that belonged in the top 32\% by quality fell from 62.8\% to 59.6\%. In
2025 it fell from 61.3\% to 57.7\%. About a third of the papers brought
onto the lists belonged there, against half of those pushed off. Actual
rebuttals can correct a mistaken review or improve a paper. The public
scores cannot tell us how often either happened.

NeurIPS measured whether agreement carries across committees. In 2021,
882 papers went through two independent committees. Their mean scores
correlated 0.575 before discussion and 0.586 after (Beygelzimer et al.
2023). The committees ended about as far apart as they began.

\section{A quarter of reviewed papers were withdrawn in
2025}\label{a-quarter-of-reviewed-papers-were-withdrawn-in-2025}

From 2023 to 2025, withdrawals rose from a fifth to a quarter of
reviewed papers. Low first-round scores predict which papers leave.

In 2023, 20.9\% of reviewed papers were withdrawn; in 2024, 21.5\%; in
2025, 25.2\%. The accepted list never shows them. For 2023 our crawls
are dated: about two thirds of that year's withdrawals came within
fifteen days of the reviews' release. In 2025 more than half of the
lowest-scoring tenth of papers (56\%) were withdrawn, and almost none of
the top tenth (0.5\%). Of the papers that left in 2024 and 2025, 84\% to
88\% came from the bottom half by score. First-round scores thus predict
both withdrawal and acceptance.

How many of the papers that left would rank highly over many independent
reviews? If authors withdrew on the basis of their first-round mean
score alone, the fitted model puts about 9\% of the 2025 withdrawals in
the top quarter by quality and 28\% in the top half. That amounts to
about 256 top-quarter papers in 2025 and 117 in 2024. The estimate
depends on what authors knew beyond their scores. Only the 2023 data
date the withdrawals (Appendix F).

\section{2027: the flood, and the
response}\label{the-flood-and-the-response}

In 2027 the flood of submissions broke far beyond the earlier trend, and
the setup changed as it arrived: a four-level scale and, in the ratings
we have, fewer reviews per paper. In the model that change cost the
accepted list far more than the rising noise did.

ICLR 2027 reviewed 40,458 papers, 115\% more than 2026's 18,791. On the
trend of 2023 to 2025, which was 54\% a year, 2027 would have had about
26,400; some 14,000, more than a third of the year, came beyond it. We
cannot say why the jump came. Easier writing with AI tools is a common
explanation, and nothing in these data tests it.

First-round noise went from 3.91 to 4.21, up 0.30 after increases of
0.28 and 0.33 in the two preceding years (Appendix G). The 2027 snapshot
leaves out papers already withdrawn. An independent archive taken at the
same time gives 4.21 for active papers and 4.14 when its 190 rated
withdrawals are included. That puts the rise at 0.23 (Appendix I).

What changed sharply was the setup. The snapshot shows 3.09 reviews per
paper by 6 November, down from 3.8 to 4.1 in the four years before, and
four rating levels where there had been six. The area-chair guide sought
three completed reviews per paper.

To see what each change cost, we rebuilt the 2026 conference and changed
the noise, the scale and the number of reviews one at a time. The share
of the accepted list that belongs fell by 5.4 points between 2026 and
2027. The rise in noise accounts for 0.7 of them. The other 4.7 come
from the setup: fewer reviews per paper cost 2.5 to 3.1 points and the
change in scale, including where its cutoffs fell, 1.6 to 2.2, depending
on which is counted first (Appendix C).

The setup costs so much because it leaves the decision to a pile of
ties. With three reviews on four levels, 21\% of the papers in the
simulated conference have exactly the mean rating where the line falls
(two 2s and a 3, for example), another 17\% lie above it, and the line
falls on that same rating for any acceptance rate from about 17\% to
38\%. The model breaks such ties at random. Ranking by the model's
estimate of quality recovers little, because what a coarse scale loses
is lost when the rating is recorded (Appendix C).

ICLR has used this setup once before. In 2020 it had a four-point scale
and, as far as we can tell, about three reviews per paper (Huang 2026).
Of the four years that Tran et al. (2020) studied, 2017 to 2020,
agreement among reviewers was lowest in 2020, and so was the chance that
an accepted paper would be accepted again. Huang found that nearly a
quarter of that year's papers got identical scores from all their
reviewers, against 5\% to 13\% in other years.

We read 2027 as what patching a strained process under load looks like.
The setup changed as the load arrived, the design of the decision stayed
as it was, and the cost fell on the papers at the line.

\section{The way out}\label{the-way-out}

Peer review at this scale is a decision process, and it can be
engineered as one. The rules for combining judgments and the order in
which they are made determine which errors reach the decision (Arbuzov
2026). A conference can say what the decision needs and work out what a
design costs and where it places papers before adopting it.

A conference must choose about three papers in ten out of tens of
thousands, from noisy readings, and every review spends researchers'
years. Today's process spends three or four reviews on every paper and
cannot separate the papers in the middle.

Going back does not help: return to 2026 and about 64\% of the accepted
list belongs, so more than a third of it did not. Go back as far as
2023, with under 5,000 papers, and 67.5\% belonged, and a rerun returned
only 56\% of the list. The process was already poor at the line before
2027.

More reviews help little. Giving every 2027 paper the 3.9 reviews that
2026 papers had would take about 33,000 more reviews: some fifty working
years on top of the 188 of the first round as it was. It would win back
about two and a half of the five points that 2027 lost: the share of the
list that belongs would rise from 58.5\% to 60.8\%.

Caps on submissions and compulsory reviewing treat the load. ICLR 2027
capped submissions at twenty per author and limited each author to one
submission without an eligible reciprocal reviewer (ICLR 2027 Program
Chairs 2026; ICLR 2026b). They change how many papers arrive and who
reviews them; neither changes how a decision is made from the reviews,
and even with 2023's small pool that decision was poor at the line. Our
earlier paper argues the same of the limits on submission now spreading
across conferences (Arbuzov et al. 2026).

In the staged funnel of our earlier paper (Arbuzov et al. 2026), an
automated gate checks form and basic soundness. Submitting authors rank
sets of short summaries of other papers. Every paper that passes the
gate gets independent judgments from twenty readers before the leading
papers go on to full reviews. The earlier paper counted the authors'
sorting time and allowed two hours for a full review. On those figures
the funnel takes 57\% as much expert time as giving every paper three
full reviews. At our three hours per full review, the funnel would take
about 87 working years for the 2027 pool, against the estimated 188
years for the first-round reviews posted by 6 November (Appendix H).

We reran the funnel with the noise of its peer judgments raised to ICLR
2027's level. When each design accepts a quarter of the papers, the
funnel recovers 76\% of the best quarter, compared with 57\% for three
full reviews scored on a continuous scale at the same noise. Using
ICLR's four-level scale and its observed review counts gives 53\%. The
largest assumption is that an author sorting short summaries judges
about as well as a reviewer reading the whole paper. The example shows
how much room a change of design could give.

A design also has to be measured. Two committees reviewing the same
papers through rebuttal and discussion would show what the later stages
really do, as NeurIPS has done twice. ICLR could run it. Without it, we
are inferring what a second committee would show from public ratings.

\section{What is known, and what this cannot
show}\label{what-is-known-and-what-this-cannot-show}

Others have measured agreement at ICLR, mostly on final scores or in a
year or two; the first round followed for five years through 2027 is
what we add. Three limits bound it.

Tran et al. (2020) measured agreement at ICLR for 2017 to 2020, and
Huang (2026) for 2017 to 2025 with a model of the same family as ours,
fitted to final, post-discussion scores; Huang finds no significant
trend in those scores, and we find none in the noise of final ratings
(Figure 2). Others have followed scores through rebuttal for 2024 and
2025 (Kargaran et al. 2025; Louis et al. 2026), and the gap between
confident and unconfident reviewers has been shown for one year
(Zahorodnii et al. 2025). None follows the first round for five years,
or reaches 2027. For 2023 to 2025, the years covered at or near release,
we also show the thinning of high ratings, the confident reviewers' edge
narrowing, and the replay that shows where the agreement of final scores
comes from; for 2027, what the setup cost. The fall of agreement when
papers draw together is known from grant review as range restriction
(Erosheva et al. 2021). Appendix J gives the extended related work.

Three limits qualify these results (Appendix I). ``Quality'' means what
independent reviewers of a paper share, and a shared judgment may be
wrong. We observe ratings, score changes, withdrawals and decisions. To
estimate where papers of a given quality land, we simulate selection on
first-round scores. The timing also differs across sources. 2023 was
crawled in release week, 2024 and 2025 were archived at release, 2026
uses the restored scores and 2027 an early-discussion snapshot.

\section{What an acceptance says now}\label{what-an-acceptance-says-now}

We think an acceptance at ICLR says less about a paper than it did. Even
for a paper in the top tenth, the risk of rejection in our model rose
from one in eight to one in four. For a paper near the line, acceptance
is close to a toss-up. Rebuttal and discussion bring scores into line,
often because a reviewer who started below the others comes up. When we
replayed that alignment without adding new information about the papers,
the list got worse. A hiring committee, a funder or a promotion panel
that reads an acceptance as a judgment of the work is reading, in part,
a draw. The bill meanwhile has grown with the pool, to 188 working years
for the first round of 2027 alone. Giving each paper as many reviews as
in 2026 would add another fifty working years and improve the list only
a little. That is too much to ask of our colleagues. The whole process
needs rethinking.

There is more at stake if nothing changes, and it is something we
believe and cannot show. A weaker badge may draw weaker reviewing, which
loosens the sorting further. And the papers withdrawn or unluckily
rejected come back as new submissions and feed the flood, whatever
started it; a model of such a loop exists (Bergstrom and Gross 2026). If
both hold, the loosening feeds itself, which is why patching cannot keep
up and why the rethinking is worth starting now. Public ratings cannot
test either belief; that takes records of who reviews and how well they
sort papers, year by year, and of where withdrawn and rejected papers
are submitted next. What remains is that, and the two-committee
measurement of what the later stages do, before the design work they
would guide.

\section*{References}\label{references}
\addcontentsline{toc}{section}{References}

\protect\phantomsection\label{refs}

\appendix

\section{Appendix A. The model and how it is
fitted}\label{appendix-a.-the-model-and-how-it-is-fitted}

\textbf{The model.} It is the model of our earlier paper (Arbuzov et al.
2026, Appendix A), with a rating scale added. A paper has a quality
\(s\), drawn from a standard normal distribution across papers. Each
review of it sees \(x = s + e\), where the error \(e\) is normal with
variance \(v\), drawn afresh for each review. The review form records
\(x\) on \(K\) levels, cut at \(K-1\) thresholds. We call \(v\) the
review noise, in multiples of the variance of quality. Two readings of
the same paper correlate \(1/(1+v)\).

\textbf{The fit.} The noise and the thresholds are fitted together by
maximum likelihood over all papers. Each paper enters through the counts
of its reviews at each level, with its unknown quality integrated out by
Gauss--Hermite quadrature. Intervals come from refitting resamples of
the papers. No reviewer identities and no review text are needed. ICLR
2027 was fitted with 160 quadrature nodes, the other years with 120.

\textbf{What the noise means.} It is a ratio: how far a paper's reviews
differ from each other against how far papers differ from one another.
It assumes one noise for every paper and every reviewer, and normal
quality and noise. What ``quality'' means, and what the fit cannot say
about it, is in Appendix I.

\textbf{ICLR 2027.} There are 40,458 papers and 125,048 reviews on four
levels, and 36,469 papers have exactly three reviews. The noise is 4.21
(95\% interval 4.00 to 4.40, from 300 resamples), with thresholds at
\(-0.88\), \(0.72\) and \(2.16\) on the standardised reading
\(z = x/\sqrt{1+v}\), which has variance 1. Two readings of a paper
correlate 0.19 (0.16 on the four recorded ratings), and one review
correlates 0.44 with quality. The means of two panels of three reviews
would correlate 0.42 (0.36 on recorded ratings). Fitted one threshold at
a time, the noise is 3.26 (1 against 2 to 4), 4.35 (1 and 2 against 3
and 4) and 3.46 (1 to 3 against 4); the middle threshold is the one
nearest the accept line.

\textbf{Does the fit reproduce the ratings?} For the papers with three
reviews, Figure 5 compares the observed counts at each mean rating with
the fitted model and with reviews unrelated to the paper (the same
overall mix of ratings, no link to the paper). Over the ten possible
mean ratings, chi-square is 48 for the model and 2,261 for unrelated
reviews. Papers with three 1s number 764 observed, 632 fitted and 247 if
unrelated; papers with three 4s number 2 observed, 2.1 fitted and 0.13
if unrelated.

At the acceptance cutoff, three ratings can sum to seven in three ways.
There were 7,667 papers rated \((2,2,3)\) against 7,719 fitted, 675
rated \((1,3,3)\) against 644 fitted, and 215 rated \((1,2,4)\) against
158 fitted. The fit is within 1\% for the most common pattern but
undercounts the widest split. Using the middle threshold's noise of 4.35
in place of 4.21, with the rating cutoffs and review counts fixed,
lowers the share of the accepted list that belongs from 58.48\% to
58.17\% at the 32\% quota.

\begin{figure}
\centering
\includegraphics[width=0.8\linewidth,height=\textheight,keepaspectratio,alt={ICLR 2027, papers with three reviews, by mean rating (log scale): observed (bars), the fitted model (dots), and reviews unrelated to the paper (crosses).}]{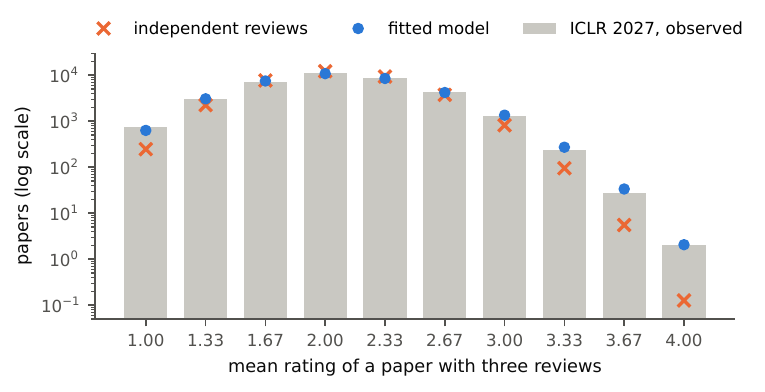}
\caption{ICLR 2027, papers with three reviews, by mean rating (log
scale): observed (bars), the fitted model (dots), and reviews unrelated
to the paper (crosses).}
\end{figure}

Table 2 checks features of the ratings that the fit does not target
directly. For each year we fitted the model, simulated the same number
of papers with the same numbers of reviews twelve times, and compared
statistics of the simulated and the real ratings. Ratings are compared
as level positions, so years with different scales are treated alike. A
lone low review is two or more levels under the rest, with the rest
within one level of each other; a lone high review is the same the other
way.

\begin{longtable}[]{@{}
  >{\raggedright\arraybackslash}p{(\linewidth - 8\tabcolsep) * \real{0.4902}}
  >{\centering\arraybackslash}p{(\linewidth - 8\tabcolsep) * \real{0.1275}}
  >{\centering\arraybackslash}p{(\linewidth - 8\tabcolsep) * \real{0.1275}}
  >{\centering\arraybackslash}p{(\linewidth - 8\tabcolsep) * \real{0.1275}}
  >{\centering\arraybackslash}p{(\linewidth - 8\tabcolsep) * \real{0.1275}}@{}}
\caption{Observed / model. The correlations in the first row are on the
recorded ratings, so they are lower than the model's correlations in the
body.}\tabularnewline
\toprule\noalign{}
\begin{minipage}[b]{\linewidth}\raggedright
\end{minipage} & \begin{minipage}[b]{\linewidth}\centering
2023
\end{minipage} & \begin{minipage}[b]{\linewidth}\centering
2024
\end{minipage} & \begin{minipage}[b]{\linewidth}\centering
2025
\end{minipage} & \begin{minipage}[b]{\linewidth}\centering
2027
\end{minipage} \\
\midrule\noalign{}
\endfirsthead
\toprule\noalign{}
\begin{minipage}[b]{\linewidth}\raggedright
\end{minipage} & \begin{minipage}[b]{\linewidth}\centering
2023
\end{minipage} & \begin{minipage}[b]{\linewidth}\centering
2024
\end{minipage} & \begin{minipage}[b]{\linewidth}\centering
2025
\end{minipage} & \begin{minipage}[b]{\linewidth}\centering
2027
\end{minipage} \\
\midrule\noalign{}
\endhead
\bottomrule\noalign{}
\endlastfoot
Two reviews of one paper, correlation & 0.245 / 0.245 & 0.204 / 0.209 &
0.193 / 0.196 & 0.156 / 0.157 \\
Four-review papers: mean of two reviews against the other two & 0.388 /
0.394 & 0.347 / 0.350 & 0.312 / 0.329 & 0.258 / 0.273 \\
Papers with one lone low review & 8.4\% / 7.7\% & 6.3\% / 6.8\% & 5.2\%
/ 5.1\% & 2.1\% / 1.9\% \\
Papers with one lone high review & 10.6\% / 11.9\% & 11.8\% / 11.1\% &
9.9\% / 9.6\% & 3.5\% / 3.2\% \\
Unanimous papers & 7.4\% / 6.2\% & 6.3\% / 5.7\% & 6.0\% / 5.3\% &
23.7\% / 22.5\% \\
\end{longtable}

Lone outlying reviews are about as common as plain independent noise
predicts, so the first round needs no special account of harsh
reviewers. Two further discrepancies appear in Table 2. It
under-predicts unanimous papers by about one point in every year, and
the split-half correlation runs a little below the model in 2025 and
2027. Both suggest that noise is not quite equal for every paper, as do
the per-threshold fits above. These are checks of fit and not of
prediction: every statistic comes from the ratings the model was fitted
to. 2026 is not checked, and nothing here tests the model against a
paper's true quality, which no one observes.

\section{Appendix B. The data}\label{appendix-b.-the-data}

\begin{longtable}[]{@{}
  >{\raggedright\arraybackslash}p{(\linewidth - 4\tabcolsep) * \real{0.0588}}
  >{\raggedright\arraybackslash}p{(\linewidth - 4\tabcolsep) * \real{0.2824}}
  >{\raggedright\arraybackslash}p{(\linewidth - 4\tabcolsep) * \real{0.6588}}@{}}
\caption{Where each year's ratings come from.}\tabularnewline
\toprule\noalign{}
\begin{minipage}[b]{\linewidth}\raggedright
ICLR
\end{minipage} & \begin{minipage}[b]{\linewidth}\raggedright
Source
\end{minipage} & \begin{minipage}[b]{\linewidth}\raggedright
What it is
\end{minipage} \\
\midrule\noalign{}
\endfirsthead
\toprule\noalign{}
\begin{minipage}[b]{\linewidth}\raggedright
ICLR
\end{minipage} & \begin{minipage}[b]{\linewidth}\raggedright
Source
\end{minipage} & \begin{minipage}[b]{\linewidth}\raggedright
What it is
\end{minipage} \\
\midrule\noalign{}
\endhead
\bottomrule\noalign{}
\endlastfoot
2023 & OpenReview crawl ({fedebotu} 2022) & Crawl of 5 November 2022,
the week reviews were released; further crawls of 20 November 2022 and
10 February 2023 (final) \\
2024, 2025 & Insights data (Paper Copilot 2025), analysed in (Kargaran
et al. 2025) & Each review's rating archived when reviews were released,
paired with its final rating (CC BY 4.0) \\
2024, 2025, decisions & Paper lists (Paper Copilot 2026) & Final status
of each paper \\
2026 & Paper lists (Paper Copilot 2026); (ICLR 2026 Program Chairs 2025)
& Public ratings after ICLR restored reviews and scores to their state
before discussion, following a security incident \\
2027 & ({shiqiangw} 2026) & Public ratings as of 6 November 2026 \\
2018 to 2025, final & Paper lists (Paper Copilot 2026) & Final
ratings \\
\end{longtable}

\textbf{One counting rule.} We count rated papers, including rated
withdrawals where available, and exclude desk rejects. The 2027 source
also leaves out 492 papers already withdrawn at the snapshot (and 1,682
desk-rejected, by the source's own counts), which is the one departure
from the rule; what it does to the 2027 noise is in Appendix I. Leaving
withdrawn papers out raises a year's fitted noise a great deal (Appendix
F), which is why one rule is used throughout. The rebuttal and alignment
figures of Appendix E, which follow papers to a decision, use papers
that were not withdrawn.

\textbf{When the scores were recorded.} The 2023 crawl was taken in
release week. The 2024 and 2025 archive captured scores at release,
although some reviewers edited them within the first hour (Kargaran et
al. 2025). For 2026 we use the paper-list record after ICLR restored
scores to their pre-discussion state (ICLR 2026 Program Chairs 2025).
For 2027 the 6 November snapshot was taken early in discussion. The 2023
noise estimate fell from 2.73 on 5 November 2022 to 2.22 on 20 November
and 1.31 in the final ratings. The timing effect in 2027 is unknown. The
final-rating series ends in 2025.

\textbf{Rating scales.} The scale is 1, 3, 5, 6, 8, 10 in 2023 to 2025;
0, 2, 4, 6, 8, 10 in 2026; and 1 to 4 in 2027. The fit estimates each
year's thresholds, which allows for the number and placement of levels
but not for reviewers using a new scale differently. We checked the
first part by refitting 2023, 2024 and 2025 after collapsing their six
levels to four (three different groupings), to three and to two. The
fitted noise moves by at most about 0.18 (2023: 2.68 to 2.83 against
2.73; 2024: 3.20 to 3.41 against 3.31; 2025: 3.53 to 3.76 against 3.58).
A coarser scale does not by itself raise the fitted noise.

\textbf{Reviews per paper.} They number about 3.8 to 4.1 in 2023 to 2026
and 3.1 in the 2027 snapshot. Whether more reviews arrived after 6
November 2026 is not known.

\textbf{ICLR 2023 paper ids.} In the final crawl the paper ids and
titles do not line up with the ratings (none of 3,432 papers with
identical review lengths at two dates carry the same id). Fits that use
only rating patterns are unaffected. Comparisons of the same paper
across dates first reserve unique exact review-length signatures, then
match remaining papers by shared lengths. This yields 4,587 matches, of
which 3,571 have a recorded decision. Within matched papers with equal
review counts, we paired the individual ratings by list position. On
3,316 of the 3,453 such decided papers, review lengths also let us
identify each pair, allowing at most one pair per paper to be assigned
by elimination. Both methods identified the same 2,270 changes among
those 12,502 ratings.

\textbf{ICLR 2024 and 2025 pairing.} The source marks how confidently
each paper's reviewers were traced across dates; we use all papers with
paired ratings regardless of that mark. Fewer than 5\% of papers lack
first-round ratings.

No data here contain reviewer identities, review text, or any measure of
a paper's quality other than its ratings and decision.

\section{Appendix C. Where papers land, by
year}\label{appendix-c.-where-papers-land-by-year}

\textbf{The simulated conference.} We draw papers with qualities from
the model, draw each paper's review count from that year's observed
distribution, draw ratings from the year's fitted noise and thresholds
on the year's own scale, rank papers by mean rating, break ties at
random, and accept the top share. We rerun with fresh reviewers to see
how much of the list returns. Each year is twenty runs of 40,000 papers.
``Belongs'' means being in the top share by quality. Perfect continuous
readings recover 100\%. With zero reviewer error but the four 2027
rating levels and their fitted standardized cutoffs, ties limit recovery
to 77.5\% at the 32\% quota. Choosing papers at random gives 32\%. We
accept 32\% of papers because ICLR accepted 31.3\% of reviewed papers in
2024 and 32.2\% in 2025.

For 2027, the share of the accepted list that belongs had a standard
deviation of 0.31 percentage points across the twenty runs. Its mean has
a Monte Carlo standard error of 0.07 points. These runs hold the fitted
model fixed. The bootstrap intervals in Figure 2 measure how the fitted
noise varies when papers are resampled.

\begin{longtable}[]{@{}
  >{\raggedright\arraybackslash}p{(\linewidth - 10\tabcolsep) * \real{0.1429}}
  >{\raggedleft\arraybackslash}p{(\linewidth - 10\tabcolsep) * \real{0.1714}}
  >{\raggedleft\arraybackslash}p{(\linewidth - 10\tabcolsep) * \real{0.1714}}
  >{\raggedleft\arraybackslash}p{(\linewidth - 10\tabcolsep) * \real{0.1714}}
  >{\raggedleft\arraybackslash}p{(\linewidth - 10\tabcolsep) * \real{0.1714}}
  >{\raggedleft\arraybackslash}p{(\linewidth - 10\tabcolsep) * \real{0.1714}}@{}}
\caption{The accepted list at each year's noise, 32\%
accepted.}\tabularnewline
\toprule\noalign{}
\begin{minipage}[b]{\linewidth}\raggedright
At the noise of ICLR
\end{minipage} & \begin{minipage}[b]{\linewidth}\raggedleft
Accepted papers that belong
\end{minipage} & \begin{minipage}[b]{\linewidth}\raggedleft
A rerun repeats
\end{minipage} & \begin{minipage}[b]{\linewidth}\raggedleft
Accepted papers from the bottom half
\end{minipage} & \begin{minipage}[b]{\linewidth}\raggedleft
A paper right at the line is accepted
\end{minipage} & \begin{minipage}[b]{\linewidth}\raggedleft
A top-tenth paper is accepted
\end{minipage} \\
\midrule\noalign{}
\endfirsthead
\toprule\noalign{}
\begin{minipage}[b]{\linewidth}\raggedright
At the noise of ICLR
\end{minipage} & \begin{minipage}[b]{\linewidth}\raggedleft
Accepted papers that belong
\end{minipage} & \begin{minipage}[b]{\linewidth}\raggedleft
A rerun repeats
\end{minipage} & \begin{minipage}[b]{\linewidth}\raggedleft
Accepted papers from the bottom half
\end{minipage} & \begin{minipage}[b]{\linewidth}\raggedleft
A paper right at the line is accepted
\end{minipage} & \begin{minipage}[b]{\linewidth}\raggedleft
A top-tenth paper is accepted
\end{minipage} \\
\midrule\noalign{}
\endhead
\bottomrule\noalign{}
\endlastfoot
2023 & 67.5\% & 56.4\% & 14.0\% & 43\% & 87\% \\
2024 & 65.7\% & 54.3\% & 15.6\% & 43\% & 85\% \\
2025 & 64.8\% & 53.6\% & 16.5\% & 42\% & 84\% \\
2026 & 63.9\% & 52.4\% & 17.3\% & 42\% & 83\% \\
2027 & 58.5\% & 47.0\% & 22.3\% & 41\% & 75\% \\
\end{longtable}

\textbf{One simulated ICLR 2027} (Figure 1). Of 40,458 papers, 18.8\%
are in the top 32\% by quality and accepted, 13.2\% in the top 32\% and
rejected, 13.2\% outside it and accepted, and 54.8\% outside it and
rejected. The median accepted paper is at the 75th percentile of quality
and the median rejected paper at the 38th. Mean rating and quality
correlate at 0.60. The mean rating where the line falls is 2.33; 21\% of
papers have exactly that mean and 17\% are above it, and the line falls
on that same value for any acceptance rate from about 17\% to 38\%.
Figure 6 shows a paper's expected mean rating by quality and its chance
of acceptance by quality. The expected mean reaches the line at the 84th
percentile of quality, and the line lies within one standard deviation
of a paper's expected mean from the 38th to the 99th percentile.

\begin{figure}
\centering
\includegraphics[width=0.92\linewidth,height=\textheight,keepaspectratio,alt={Left: a paper's expected mean of three ratings by its quality percentile, with one standard deviation, in the simulated ICLR 2027; the accept line at 32\% is 2.33. Right: a paper's chance of acceptance by quality at ICLR 2027's noise (4.21), at the NeurIPS noise of our earlier paper (2.45), and on the same four levels with zero reviewer error and uniform ties.}]{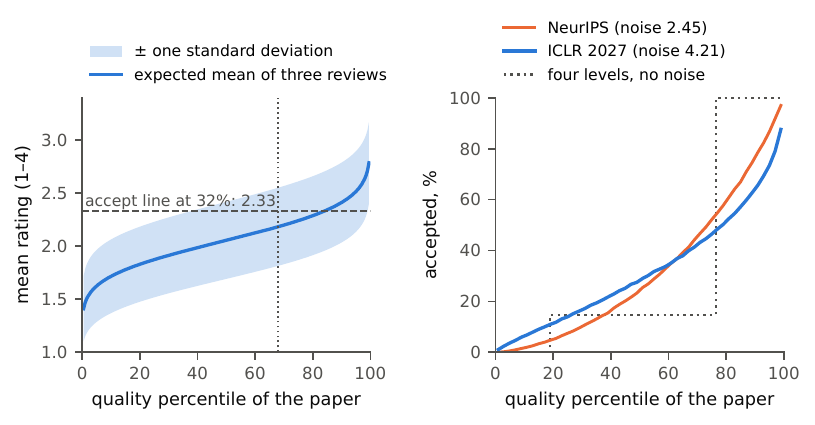}
\caption{Left: a paper's expected mean of three ratings by its quality
percentile, with one standard deviation, in the simulated ICLR 2027; the
accept line at 32\% is 2.33. Right: a paper's chance of acceptance by
quality at ICLR 2027's noise (4.21), at the NeurIPS noise of our earlier
paper (2.45), and on the same four levels with zero reviewer error and
uniform ties.}
\end{figure}

\textbf{What the 2027 step is made of.} We rebuilt the 2027 conference
from 2026's, one change at a time, on 2026's real six-level scale
wherever the scale is held, 32\% accepted, thirty runs each. Changing
the scale changes the number and values of its categories and the fitted
positions of the cutoffs together.

\begin{longtable}[]{@{}
  >{\raggedright\arraybackslash}p{(\linewidth - 8\tabcolsep) * \real{0.6522}}
  >{\raggedleft\arraybackslash}p{(\linewidth - 8\tabcolsep) * \real{0.0870}}
  >{\raggedleft\arraybackslash}p{(\linewidth - 8\tabcolsep) * \real{0.0870}}
  >{\raggedleft\arraybackslash}p{(\linewidth - 8\tabcolsep) * \real{0.0870}}
  >{\raggedleft\arraybackslash}p{(\linewidth - 8\tabcolsep) * \real{0.0870}}@{}}
\caption{The 2027 step taken apart. Thirty runs each; the table above
gives 63.9\% and 58.5\% for the first and fifth settings, from twenty
runs.}\tabularnewline
\toprule\noalign{}
\begin{minipage}[b]{\linewidth}\raggedright
Setting
\end{minipage} & \begin{minipage}[b]{\linewidth}\raggedleft
Belongs
\end{minipage} & \begin{minipage}[b]{\linewidth}\raggedleft
A rerun repeats
\end{minipage} & \begin{minipage}[b]{\linewidth}\raggedleft
From the bottom half
\end{minipage} & \begin{minipage}[b]{\linewidth}\raggedleft
Top-tenth paper accepted
\end{minipage} \\
\midrule\noalign{}
\endfirsthead
\toprule\noalign{}
\begin{minipage}[b]{\linewidth}\raggedright
Setting
\end{minipage} & \begin{minipage}[b]{\linewidth}\raggedleft
Belongs
\end{minipage} & \begin{minipage}[b]{\linewidth}\raggedleft
A rerun repeats
\end{minipage} & \begin{minipage}[b]{\linewidth}\raggedleft
From the bottom half
\end{minipage} & \begin{minipage}[b]{\linewidth}\raggedleft
Top-tenth paper accepted
\end{minipage} \\
\midrule\noalign{}
\endhead
\bottomrule\noalign{}
\endlastfoot
2026 as it was (2026 noise, six levels, 2026's reviews per paper) &
63.8\% & 52.4\% & 17.1\% & 82.7\% \\
Only the noise changed to 2027's & 63.1\% & 51.6\% & 17.9\% & 81.5\% \\
2027 noise and 2027's reviews per paper, six levels & 60.0\% & 48.3\% &
20.9\% & 77.5\% \\
2027 noise and four levels, 2026's reviews per paper & 60.9\% & 49.5\% &
20.1\% & 78.6\% \\
2027 as in the snapshot (all three) & 58.4\% & 46.9\% & 22.3\% &
74.7\% \\
2026 noise on four levels, 2026's reviews per paper & 61.7\% & 50.4\% &
19.4\% & 79.9\% \\
\end{longtable}

Of the 5.4 points lost between 2026 and 2027, the rise in noise accounts
for 0.7 (63.8\% to 63.1\%, scale and reviews held at 2026's). The other
4.7 are the setup, and how they divide depends on the order. Counting
fewer reviews per paper first, they cost 3.1 points (63.1\% to 60.0\%,
on six levels) and the four-level scale then 1.6 (60.0\% to 58.4\%).
Counting the scale first, it costs 2.2 (63.1\% to 60.9\%, at 2026's
reviews per paper) and fewer reviews then 2.5 (60.9\% to 58.4\%). The
four-level scale alone, with 2026's noise and reviews, costs 2.1 (63.8\%
to 61.7\%).

\textbf{At 28\% accepted.} ICLR 2026 accepted nearer 28\% than 32\%
(Appendix I). The same conference at 28\%, twenty runs each:

\begin{longtable}[]{@{}
  >{\raggedright\arraybackslash}p{(\linewidth - 8\tabcolsep) * \real{0.1724}}
  >{\raggedleft\arraybackslash}p{(\linewidth - 8\tabcolsep) * \real{0.2069}}
  >{\raggedleft\arraybackslash}p{(\linewidth - 8\tabcolsep) * \real{0.2069}}
  >{\raggedleft\arraybackslash}p{(\linewidth - 8\tabcolsep) * \real{0.2069}}
  >{\raggedleft\arraybackslash}p{(\linewidth - 8\tabcolsep) * \real{0.2069}}@{}}
\caption{The accepted list at each year's noise, 28\%
accepted.}\tabularnewline
\toprule\noalign{}
\begin{minipage}[b]{\linewidth}\raggedright
At the noise of ICLR
\end{minipage} & \begin{minipage}[b]{\linewidth}\raggedleft
Belongs
\end{minipage} & \begin{minipage}[b]{\linewidth}\raggedleft
A rerun repeats
\end{minipage} & \begin{minipage}[b]{\linewidth}\raggedleft
From the bottom half
\end{minipage} & \begin{minipage}[b]{\linewidth}\raggedleft
Top-tenth paper accepted
\end{minipage} \\
\midrule\noalign{}
\endfirsthead
\toprule\noalign{}
\begin{minipage}[b]{\linewidth}\raggedright
At the noise of ICLR
\end{minipage} & \begin{minipage}[b]{\linewidth}\raggedleft
Belongs
\end{minipage} & \begin{minipage}[b]{\linewidth}\raggedleft
A rerun repeats
\end{minipage} & \begin{minipage}[b]{\linewidth}\raggedleft
From the bottom half
\end{minipage} & \begin{minipage}[b]{\linewidth}\raggedleft
Top-tenth paper accepted
\end{minipage} \\
\midrule\noalign{}
\endhead
\bottomrule\noalign{}
\endlastfoot
2023 & 65.1\% & 53.5\% & 12.1\% & 83.7\% \\
2024 & 62.9\% & 50.9\% & 13.7\% & 81.1\% \\
2025 & 62.8\% & 50.7\% & 14.0\% & 80.4\% \\
2026 & 61.0\% & 48.9\% & 15.5\% & 78.0\% \\
2027 & 55.1\% & 42.9\% & 20.9\% & 69.9\% \\
\end{longtable}

It is the same picture, a little harsher: a top-tenth paper is rejected
three times in ten at 2027's setup.

\textbf{The ends of the noise interval.} Running the conference at the
ends of each fitted noise's 95\% interval, 32\% accepted, moves the
figures by about one point either way:

\begin{longtable}[]{@{}
  >{\raggedright\arraybackslash}p{(\linewidth - 10\tabcolsep) * \real{0.1429}}
  >{\raggedleft\arraybackslash}p{(\linewidth - 10\tabcolsep) * \real{0.1714}}
  >{\raggedleft\arraybackslash}p{(\linewidth - 10\tabcolsep) * \real{0.1714}}
  >{\raggedleft\arraybackslash}p{(\linewidth - 10\tabcolsep) * \real{0.1714}}
  >{\raggedleft\arraybackslash}p{(\linewidth - 10\tabcolsep) * \real{0.1714}}
  >{\raggedleft\arraybackslash}p{(\linewidth - 10\tabcolsep) * \real{0.1714}}@{}}
\caption{The accepted list at the ends of the noise interval (the
60-resample interval; the 300-resample intervals of Appendix G are
within 0.1 of these ends).}\tabularnewline
\toprule\noalign{}
\begin{minipage}[b]{\linewidth}\raggedright
\end{minipage} & \begin{minipage}[b]{\linewidth}\raggedleft
Noise
\end{minipage} & \begin{minipage}[b]{\linewidth}\raggedleft
Belongs
\end{minipage} & \begin{minipage}[b]{\linewidth}\raggedleft
A rerun repeats
\end{minipage} & \begin{minipage}[b]{\linewidth}\raggedleft
From the bottom half
\end{minipage} & \begin{minipage}[b]{\linewidth}\raggedleft
Top-tenth paper accepted
\end{minipage} \\
\midrule\noalign{}
\endfirsthead
\toprule\noalign{}
\begin{minipage}[b]{\linewidth}\raggedright
\end{minipage} & \begin{minipage}[b]{\linewidth}\raggedleft
Noise
\end{minipage} & \begin{minipage}[b]{\linewidth}\raggedleft
Belongs
\end{minipage} & \begin{minipage}[b]{\linewidth}\raggedleft
A rerun repeats
\end{minipage} & \begin{minipage}[b]{\linewidth}\raggedleft
From the bottom half
\end{minipage} & \begin{minipage}[b]{\linewidth}\raggedleft
Top-tenth paper accepted
\end{minipage} \\
\midrule\noalign{}
\endhead
\bottomrule\noalign{}
\endlastfoot
2023, low end & 2.50 & 68.3\% & 57.6\% & 13.1\% & 88.3\% \\
2023, fitted & 2.73 & 67.3\% & 56.4\% & 14.1\% & 87.1\% \\
2023, high end & 2.97 & 66.5\% & 55.3\% & 14.9\% & 86.0\% \\
2027, low end & 4.07 & 58.9\% & 47.4\% & 22.0\% & 75.6\% \\
2027, fitted & 4.21 & 58.4\% & 47.1\% & 22.3\% & 74.9\% \\
2027, high end & 4.48 & 57.8\% & 46.4\% & 23.0\% & 73.9\% \\
\end{longtable}

\textbf{A better ranking rule.} We also ranked papers by the model's
best estimate of a paper's quality given its ratings, in place of the
mean rating with ties broken at random. The share of the accepted list
that belongs is 63.8\% against 63.9\% for 2026 and 58.7\% against 58.5\%
for 2027, so ranking by posterior mean changes little. With continuous
scores in place of the recorded ratings it would be 65.1\% and 61.7\%.
The six-level scale loses 1.3 points in 2026 and the four-level scale
3.2 in 2027, and what they lose is lost when the rating is recorded, not
when ratings are averaged or ties broken.

\textbf{Limits.} This is a conference that decides on first-round means
with ties broken at random. ICLR decides after rebuttal and discussion,
through area chairs; Appendix E gives how far the first round tells
ICLR's decisions. The model's small misfits (Appendix A) carry over.

\section{Appendix D. What changed in the
ratings}\label{appendix-d.-what-changed-in-the-ratings}

ICLR 2023 to 2025 shared one scale (1, 3, 5, 6, 8, 10) and one form, so
rating points can be compared directly. For the variance estimates we
give each paper with at least two ratings equal weight. If paper \(i\)
has \(n_i\) ratings with sample variance \(s_i^2\), the within-paper
variance \(W\) is the average of \(s_i^2\) across papers. The
between-paper variance \(B\) is the sample variance of the paper means
minus the average of \(s_i^2/n_i\), the part due to having only a few
reviews. Both are in squared rating points. The estimated correlation of
two recorded ratings of one paper is \(B/(B+W)\).

\begin{longtable}[]{@{}
  >{\raggedright\arraybackslash}p{(\linewidth - 12\tabcolsep) * \real{0.1220}}
  >{\raggedleft\arraybackslash}p{(\linewidth - 12\tabcolsep) * \real{0.1463}}
  >{\raggedleft\arraybackslash}p{(\linewidth - 12\tabcolsep) * \real{0.1463}}
  >{\raggedleft\arraybackslash}p{(\linewidth - 12\tabcolsep) * \real{0.1463}}
  >{\raggedleft\arraybackslash}p{(\linewidth - 12\tabcolsep) * \real{0.1463}}
  >{\raggedleft\arraybackslash}p{(\linewidth - 12\tabcolsep) * \real{0.1463}}
  >{\raggedleft\arraybackslash}p{(\linewidth - 12\tabcolsep) * \real{0.1463}}@{}}
\caption{Spread among and between papers.}\tabularnewline
\toprule\noalign{}
\begin{minipage}[b]{\linewidth}\raggedright
ICLR
\end{minipage} & \begin{minipage}[b]{\linewidth}\raggedleft
Spread among one paper's reviews
\end{minipage} & \begin{minipage}[b]{\linewidth}\raggedleft
Spread between papers
\end{minipage} & \begin{minipage}[b]{\linewidth}\raggedleft
Ratio
\end{minipage} & \begin{minipage}[b]{\linewidth}\raggedleft
Ratings of 5 or 6
\end{minipage} & \begin{minipage}[b]{\linewidth}\raggedleft
Ratings of 8 or 10
\end{minipage} & \begin{minipage}[b]{\linewidth}\raggedleft
Mean rating
\end{minipage} \\
\midrule\noalign{}
\endfirsthead
\toprule\noalign{}
\begin{minipage}[b]{\linewidth}\raggedright
ICLR
\end{minipage} & \begin{minipage}[b]{\linewidth}\raggedleft
Spread among one paper's reviews
\end{minipage} & \begin{minipage}[b]{\linewidth}\raggedleft
Spread between papers
\end{minipage} & \begin{minipage}[b]{\linewidth}\raggedleft
Ratio
\end{minipage} & \begin{minipage}[b]{\linewidth}\raggedleft
Ratings of 5 or 6
\end{minipage} & \begin{minipage}[b]{\linewidth}\raggedleft
Ratings of 8 or 10
\end{minipage} & \begin{minipage}[b]{\linewidth}\raggedleft
Mean rating
\end{minipage} \\
\midrule\noalign{}
\endhead
\bottomrule\noalign{}
\endlastfoot
2023 & 2.27 & 0.73 & 3.10 & 56\% & 12.4\% & 4.94 \\
2024 & 2.18 & 0.57 & 3.84 & 58\% & 10.2\% & 4.87 \\
2025 & 2.05 & 0.49 & 4.15 & 59\% & 8.0\% & 4.78 \\
\end{longtable}

Reviewers of one paper differ slightly less than before. Papers differ
from one another much less: the variance between papers fell by a third
in two years, and its square root, the standard deviation, by 18\%, from
0.85 to 0.70 rating points. Ratings bunched toward the middle and high
ratings became rarer each year, so a review now says less about where a
paper stands among the others. Two readings fit: the submissions became
more alike, or reviewers became less able or less willing to tell them
apart. The ratio here is on raw ratings, so it differs a little from the
fitted noise.

\textbf{What did not change.}

\begin{longtable}[]{@{}lrrr@{}}
\caption{What stayed put.}\tabularnewline
\toprule\noalign{}
& 2023 & 2024 & 2025 \\
\midrule\noalign{}
\endfirsthead
\toprule\noalign{}
& 2023 & 2024 & 2025 \\
\midrule\noalign{}
\endhead
\bottomrule\noalign{}
\endlastfoot
Mean stated confidence & 3.62 & 3.64 & 3.66 \\
Reviews at confidence 4 or 5 & 59.2\% & 59.5\% & 60.2\% \\
Ratings of 1 or 3 & 31.3\% & 31.8\% & 32.6\% \\
Papers with a mean of 4 or less & 24.0\% & 24.8\% & 25.3\% \\
\end{longtable}

The scale and the form did not change either. The number of reviews per
paper did rise a little: 3.75, 3.86 and 4.05, with 68\%, 74\% and 84\%
of papers having four or more reviews.

\textbf{What changed: the top thinned out.}

\begin{longtable}[]{@{}
  >{\raggedright\arraybackslash}p{(\linewidth - 6\tabcolsep) * \real{0.6400}}
  >{\raggedleft\arraybackslash}p{(\linewidth - 6\tabcolsep) * \real{0.1200}}
  >{\raggedleft\arraybackslash}p{(\linewidth - 6\tabcolsep) * \real{0.1200}}
  >{\raggedleft\arraybackslash}p{(\linewidth - 6\tabcolsep) * \real{0.1200}}@{}}
\caption{The top of the ratings.}\tabularnewline
\toprule\noalign{}
\begin{minipage}[b]{\linewidth}\raggedright
\end{minipage} & \begin{minipage}[b]{\linewidth}\raggedleft
2023
\end{minipage} & \begin{minipage}[b]{\linewidth}\raggedleft
2024
\end{minipage} & \begin{minipage}[b]{\linewidth}\raggedleft
2025
\end{minipage} \\
\midrule\noalign{}
\endfirsthead
\toprule\noalign{}
\begin{minipage}[b]{\linewidth}\raggedright
\end{minipage} & \begin{minipage}[b]{\linewidth}\raggedleft
2023
\end{minipage} & \begin{minipage}[b]{\linewidth}\raggedleft
2024
\end{minipage} & \begin{minipage}[b]{\linewidth}\raggedleft
2025
\end{minipage} \\
\midrule\noalign{}
\endhead
\bottomrule\noalign{}
\endlastfoot
Ratings of 8 & 12.0\% & 9.9\% & 7.8\% \\
Ratings of 5 & 32.3\% & 33.9\% & 35.6\% \\
Ratings of 6 & 23.9\% & 24.1\% & 23.7\% \\
Papers with a mean of 6 or more & 20.7\% & 17.1\% & 13.3\% \\
90th percentile of paper means & 6.33 & 6.25 & 6.00 \\
Median paper mean & 5.00 & 4.80 & 4.75 \\
Reviewed papers & 4,833 & 6,954 & 11,423 \\
Papers with a mean of 6 or more, in number & about 1,000 & about 1,190 &
about 1,520 \\
\end{longtable}

All of the bunching is at the top; the low end of the pool is the same
share as before. The pool grew 2.4 times, and the papers that stand out
grew by half. The endpoint decline appears at every level of stated
confidence: for confidence 4, ratings of 8 or 10 fell from 12.2\% to
7.6\%, and for confidence 5, from 9.8\% to 5.4\%.

\textbf{Confident reviewers lost some of their edge.} For each
confidence group we use the same correlation and give each paper with at
least two reviews in that group equal weight. In 2023, 2024 and 2025,
respectively, 3,537, 5,220 and 8,786 papers qualified at confidence 4 or
5, and 2,359, 3,403 and 5,752 at confidence 1 to 3.

\begin{longtable}[]{@{}lrrr@{}}
\caption{Agreement by stated confidence.}\tabularnewline
\toprule\noalign{}
& 2023 & 2024 & 2025 \\
\midrule\noalign{}
\endfirsthead
\toprule\noalign{}
& 2023 & 2024 & 2025 \\
\midrule\noalign{}
\endhead
\bottomrule\noalign{}
\endlastfoot
Both reviewers at confidence 4 or 5 & 0.269 & 0.239 & 0.203 \\
Both reviewers at confidence 1 to 3 & 0.199 & 0.165 & 0.164 \\
All reviews & 0.244 & 0.207 & 0.194 \\
\end{longtable}

The same correlations as noise,
\((1-\text{correlation})/\text{correlation}\):

\begin{longtable}[]{@{}lrrr@{}}
\caption{Confident and less confident pairs, as noise.}\tabularnewline
\toprule\noalign{}
& 2023 & 2024 & 2025 \\
\midrule\noalign{}
\endfirsthead
\toprule\noalign{}
& 2023 & 2024 & 2025 \\
\midrule\noalign{}
\endhead
\bottomrule\noalign{}
\endlastfoot
Two confident reviewers & 2.72 & 3.18 & 3.93 \\
Two less confident reviewers & 4.03 & 5.06 & 5.10 \\
Ratio, less confident to confident & 1.48 & 1.59 & 1.30 \\
\end{longtable}

Confident pairs lost agreement in both years. Less confident pairs lost
it from 2023 to 2024 and then held. The gap between the two groups is
narrowest in 2025 but did not narrow in 2024, so the narrowing is a
change seen in one year and not a steady trend.

The confident reviewers' ratings of one paper did not scatter more; what
fell was the spread between the papers they rated, by more than a third,
and by more than a quarter for less confident reviewers (variances in
squared rating points, over papers with two or more such reviews):

\begin{longtable}[]{@{}
  >{\raggedright\arraybackslash}p{(\linewidth - 6\tabcolsep) * \real{0.6038}}
  >{\centering\arraybackslash}p{(\linewidth - 6\tabcolsep) * \real{0.1321}}
  >{\centering\arraybackslash}p{(\linewidth - 6\tabcolsep) * \real{0.1321}}
  >{\centering\arraybackslash}p{(\linewidth - 6\tabcolsep) * \real{0.1321}}@{}}
\caption{Spread among and between papers, by stated
confidence.}\tabularnewline
\toprule\noalign{}
\begin{minipage}[b]{\linewidth}\raggedright
\end{minipage} & \begin{minipage}[b]{\linewidth}\centering
2023
\end{minipage} & \begin{minipage}[b]{\linewidth}\centering
2024
\end{minipage} & \begin{minipage}[b]{\linewidth}\centering
2025
\end{minipage} \\
\midrule\noalign{}
\endfirsthead
\toprule\noalign{}
\begin{minipage}[b]{\linewidth}\raggedright
\end{minipage} & \begin{minipage}[b]{\linewidth}\centering
2023
\end{minipage} & \begin{minipage}[b]{\linewidth}\centering
2024
\end{minipage} & \begin{minipage}[b]{\linewidth}\centering
2025
\end{minipage} \\
\midrule\noalign{}
\endhead
\bottomrule\noalign{}
\endlastfoot
Confident: among a paper's reviews / between papers & 2.34 / 0.86 & 2.23
/ 0.70 & 2.12 / 0.54 \\
Less confident: among a paper's reviews / between papers & 2.02 / 0.50 &
1.96 / 0.39 & 1.82 / 0.36 \\
\end{longtable}

\textbf{What the ratings cannot tell.} Whether the new submissions are
mostly middling papers, so that standout work is a shrinking share, or
whether reviewers have become reluctant to give an 8 to any paper. The
steady low end argues a little against reviewers simply becoming
harsher, since that would add low ratings too, but it does not rule out
a reluctance specific to the top. Stated confidence is the reviewer's
own report, and these are three years.

\section{Appendix E. After the first
round}\label{appendix-e.-after-the-first-round}

Most first-round ratings stay put. The ratings that rise tend to start
below the other reviews, and the gain in agreement is concentrated on
papers eventually accepted. The score changes show where that gain came
from. A simulation asks how much of it would arise if reviewers changed
their scores without learning anything new about the papers.

\textbf{Data.} The papers are those not withdrawn and with a decision.
ICLR 2023 is two crawls, of 5 November 2022 and 10 February 2023,
matched by paper. For ICLR 2024 and 2025 we have each review's rating
before and after rebuttal, from the archive that Kargaran et al.~study
(Kargaran et al. 2025). They report that most scores are unchanged and
that changes are mostly increases concentrated on borderline papers. Our
2024 and 2025 rows are a second analysis of that archive, not
independent confirmation. Including all paired ratings gives 81.3\%
(21,853/26,878) and 75.5\% (34,988/46,356) unchanged, consistent with
their reported 81\% and 75\%. Restricting to decided papers gives 77.1\%
and 68.6\%. The 2023 rows use a different archive.

\textbf{How much changes, and how far the first round already decides.}
Most ratings never change, though fewer each year, and nearly all the
changes are upward. About three in four of the first-round leaders are
accepted, so the other quarter of the accepted list comes from behind.
The first-round mean separates accepted from rejected papers a little
worse each year. ``Accepted'' is the area chairs' decision, made on more
than the ratings.

\begin{longtable}[]{@{}
  >{\raggedright\arraybackslash}p{(\linewidth - 6\tabcolsep) * \real{0.4316}}
  >{\raggedleft\arraybackslash}p{(\linewidth - 6\tabcolsep) * \real{0.1895}}
  >{\raggedleft\arraybackslash}p{(\linewidth - 6\tabcolsep) * \real{0.1895}}
  >{\raggedleft\arraybackslash}p{(\linewidth - 6\tabcolsep) * \real{0.1895}}@{}}
\caption{How far ratings change and how well they predict the decision.
Separation (area under the ROC curve) is the probability that a randomly
drawn accepted paper outscores a randomly drawn rejected one, with ties
counting one half. First-round leaders are the papers leading by
first-round mean, as many as were accepted.}\tabularnewline
\toprule\noalign{}
\begin{minipage}[b]{\linewidth}\raggedright
\end{minipage} & \begin{minipage}[b]{\linewidth}\raggedleft
2023
\end{minipage} & \begin{minipage}[b]{\linewidth}\raggedleft
2024
\end{minipage} & \begin{minipage}[b]{\linewidth}\raggedleft
2025
\end{minipage} \\
\midrule\noalign{}
\endfirsthead
\toprule\noalign{}
\begin{minipage}[b]{\linewidth}\raggedright
\end{minipage} & \begin{minipage}[b]{\linewidth}\raggedleft
2023
\end{minipage} & \begin{minipage}[b]{\linewidth}\raggedleft
2024
\end{minipage} & \begin{minipage}[b]{\linewidth}\raggedleft
2025
\end{minipage} \\
\midrule\noalign{}
\endhead
\bottomrule\noalign{}
\endlastfoot
Ratings unchanged & 80.9\% & 77.1\% & 68.6\% \\
Ratings raised / lowered & 17\% / 2\% & 21.5\% / 1.4\% & 29.9\% /
1.5\% \\
Papers whose mean did not move & 52\% & 44\% & 32\% \\
Correlation of first-round and final paper means & 0.93 & 0.91 & 0.89 \\
First-round leaders accepted & 79\% & 75\% & 75\% \\
How well the first-round mean separates accepted from rejected & 0.915 &
0.877 & 0.870 \\
How well the final mean separates accepted from rejected & 0.965 & 0.941
& 0.944 \\
Papers accepted (of those not withdrawn) & 40\% & 40\% & 43\% \\
\end{longtable}

\textbf{Which ratings change, and which way.} We measure a rating's
position by subtracting the mean of the paper's other first-round
ratings from that rating. A review below the others' mean was more
likely to rise, and one above it more likely to fall. The asymmetry is
the finding: even for a rating at least 2.5 points above the others,
lowering is about one case in twenty, while for a rating more than 2.5
points below them, raising is one in two in 2025. The correlation of a
rating's change with its distance from its co-reviewers is \(-0.34\) in
both years.

\begin{longtable}[]{@{}
  >{\raggedright\arraybackslash}p{(\linewidth - 8\tabcolsep) * \real{0.3766}}
  >{\raggedleft\arraybackslash}p{(\linewidth - 8\tabcolsep) * \real{0.1558}}
  >{\raggedleft\arraybackslash}p{(\linewidth - 8\tabcolsep) * \real{0.1558}}
  >{\raggedleft\arraybackslash}p{(\linewidth - 8\tabcolsep) * \real{0.1558}}
  >{\raggedleft\arraybackslash}p{(\linewidth - 8\tabcolsep) * \real{0.1558}}@{}}
\caption{A rating's change by where it stood among its co-reviewers in
the first round. Papers not withdrawn, three or more
reviews.}\tabularnewline
\toprule\noalign{}
\begin{minipage}[b]{\linewidth}\raggedright
First-round position
\end{minipage} & \begin{minipage}[b]{\linewidth}\raggedleft
Raised, 2024
\end{minipage} & \begin{minipage}[b]{\linewidth}\raggedleft
Raised, 2025
\end{minipage} & \begin{minipage}[b]{\linewidth}\raggedleft
Lowered, 2024
\end{minipage} & \begin{minipage}[b]{\linewidth}\raggedleft
Lowered, 2025
\end{minipage} \\
\midrule\noalign{}
\endfirsthead
\toprule\noalign{}
\begin{minipage}[b]{\linewidth}\raggedright
First-round position
\end{minipage} & \begin{minipage}[b]{\linewidth}\raggedleft
Raised, 2024
\end{minipage} & \begin{minipage}[b]{\linewidth}\raggedleft
Raised, 2025
\end{minipage} & \begin{minipage}[b]{\linewidth}\raggedleft
Lowered, 2024
\end{minipage} & \begin{minipage}[b]{\linewidth}\raggedleft
Lowered, 2025
\end{minipage} \\
\midrule\noalign{}
\endhead
\bottomrule\noalign{}
\endlastfoot
More than 2.5 below & 43\% & 51\% & 0.3\% & 0.6\% \\
About 2 below & 34\% & 43\% & 0.3\% & 0.5\% \\
About 1 below & 29\% & 40\% & 0.4\% & 0.8\% \\
Level & 24\% & 35\% & 0.6\% & 1.1\% \\
About 1 above & 14\% & 22\% & 1.6\% & 1.6\% \\
About 2 above & 6\% & 10\% & 3.3\% & 2.7\% \\
At least 2.5 above & 1\% & 1.5\% & 5.8\% & 5.1\% \\
\end{longtable}

\textbf{Papers with a lone low review.} Papers with one lone low review
(two or more points under every other review, the others within two
points of each other) are about 30\% of the papers not withdrawn. The
lone low review is raised in 40\% of them in 2024 and 49\% in 2025.
These papers are accepted 32\% and 35\% of the time, against 24\% and
29\% for papers with the same first-round mean and no such review. When
the lone low review is raised the paper is accepted 54\% and 53\% of the
time; when it is not, 18\% and 19\%. The records cannot tell whether the
rise preceded the decision. At a fixed quota, each paper that enters the
list displaces another.

\textbf{Where the raises sit.} The raised ratings are on the papers that
end up accepted. The table counts every rating that changed, by what
became of its paper.

\begin{longtable}[]{@{}
  >{\raggedright\arraybackslash}p{(\linewidth - 6\tabcolsep) * \real{0.5281}}
  >{\raggedleft\arraybackslash}p{(\linewidth - 6\tabcolsep) * \real{0.1573}}
  >{\raggedleft\arraybackslash}p{(\linewidth - 6\tabcolsep) * \real{0.1573}}
  >{\raggedleft\arraybackslash}p{(\linewidth - 6\tabcolsep) * \real{0.1573}}@{}}
\caption{Numbers of ratings raised and lowered, by the paper's decision.
Papers not withdrawn.}\tabularnewline
\toprule\noalign{}
\begin{minipage}[b]{\linewidth}\raggedright
\end{minipage} & \begin{minipage}[b]{\linewidth}\raggedleft
2023
\end{minipage} & \begin{minipage}[b]{\linewidth}\raggedleft
2024
\end{minipage} & \begin{minipage}[b]{\linewidth}\raggedleft
2025
\end{minipage} \\
\midrule\noalign{}
\endfirsthead
\toprule\noalign{}
\begin{minipage}[b]{\linewidth}\raggedright
\end{minipage} & \begin{minipage}[b]{\linewidth}\raggedleft
2023
\end{minipage} & \begin{minipage}[b]{\linewidth}\raggedleft
2024
\end{minipage} & \begin{minipage}[b]{\linewidth}\raggedleft
2025
\end{minipage} \\
\midrule\noalign{}
\endhead
\bottomrule\noalign{}
\endlastfoot
Papers that end up accepted: raised / lowered & 1,532 / 22 & 2,930 / 45
& 6,658 / 103 \\
Papers that end up rejected: raised / lowered & 747 / 197 & 1,599 / 258
& 3,690 / 416 \\
\end{longtable}

On accepted papers lowering is almost absent. On rejected papers it is
rarer than raising but not absent. The same shows among papers with the
same first-round mark: in 2023, papers with a first-round mean of 5.25
to 5.75 were accepted 45\% of the time, and their mean rating rose by
0.69 points on average if they were accepted and by 0.10 if they were
rejected.

\textbf{Where the recorded agreement rose.} The ratings of one paper
agree about twice as much after the process as before it: the
correlation of two reviews of one paper goes from 0.17 to 0.33 in 2024
and from 0.15 to 0.33 in 2025. We start with the recorded first-round
ratings, replace some of them with their actual final values, and
measure agreement again using the equal-paper weighting of Appendix D.
We cut the changes two ways, by where a rating stood and by what became
of its paper.

By where a rating stood, in correlations of the recorded ratings of
papers not withdrawn. They are lower than the model's correlations in
Table 1, which count withdrawn papers and allow for the coarseness of
the scale.

\begin{longtable}[]{@{}
  >{\raggedright\arraybackslash}p{(\linewidth - 4\tabcolsep) * \real{0.6883}}
  >{\raggedleft\arraybackslash}p{(\linewidth - 4\tabcolsep) * \real{0.1558}}
  >{\raggedleft\arraybackslash}p{(\linewidth - 4\tabcolsep) * \real{0.1558}}@{}}
\caption{Agreement when only some ratings take their final value; the
others keep their first-round rating.}\tabularnewline
\toprule\noalign{}
\begin{minipage}[b]{\linewidth}\raggedright
Whose final rating is used
\end{minipage} & \begin{minipage}[b]{\linewidth}\raggedleft
2024
\end{minipage} & \begin{minipage}[b]{\linewidth}\raggedleft
2025
\end{minipage} \\
\midrule\noalign{}
\endfirsthead
\toprule\noalign{}
\begin{minipage}[b]{\linewidth}\raggedright
Whose final rating is used
\end{minipage} & \begin{minipage}[b]{\linewidth}\raggedleft
2024
\end{minipage} & \begin{minipage}[b]{\linewidth}\raggedleft
2025
\end{minipage} \\
\midrule\noalign{}
\endhead
\bottomrule\noalign{}
\endlastfoot
None (first round) & 0.17 & 0.15 \\
Only ratings 1.5+ points below their co-reviewers & 0.30 & 0.29 \\
Only ratings level with or above their co-reviewers & 0.16 & 0.14 \\
All (final) & 0.33 & 0.33 \\
\end{longtable}

The ratings that stood 1.5 or more points below their co-reviewers are
about 18\% of all reviews, and letting only them change gives most of
the gain. The gain is mostly papers moving apart: the spread between
papers goes from 0.46 to 0.93 in 2024 and from 0.37 to 0.90 in 2025,
while the spread among a paper's reviews goes from 2.23 to 1.87 and from
2.12 to 1.81 (variances in squared rating points, on papers not
withdrawn, so not the figures of Appendix D).

By what became of the paper, in noise, the ratio of the spread among a
paper's reviews to the spread between papers, where less noise is more
agreement. For 2024 and 2025 it is the fitted noise of Appendix A on
papers not withdrawn; for 2023 it is the raw ratio of the two spreads on
the matched papers not withdrawn.

\begin{longtable}[]{@{}
  >{\raggedright\arraybackslash}p{(\linewidth - 6\tabcolsep) * \real{0.6386}}
  >{\raggedleft\arraybackslash}p{(\linewidth - 6\tabcolsep) * \real{0.1205}}
  >{\raggedleft\arraybackslash}p{(\linewidth - 6\tabcolsep) * \real{0.1205}}
  >{\raggedleft\arraybackslash}p{(\linewidth - 6\tabcolsep) * \real{0.1205}}@{}}
\caption{Noise when only some ratings take their final value, by the
paper's decision; the others keep their first-round
rating.}\tabularnewline
\toprule\noalign{}
\begin{minipage}[b]{\linewidth}\raggedright
Whose final rating is used
\end{minipage} & \begin{minipage}[b]{\linewidth}\raggedleft
2023
\end{minipage} & \begin{minipage}[b]{\linewidth}\raggedleft
2024
\end{minipage} & \begin{minipage}[b]{\linewidth}\raggedleft
2025
\end{minipage} \\
\midrule\noalign{}
\endfirsthead
\toprule\noalign{}
\begin{minipage}[b]{\linewidth}\raggedright
Whose final rating is used
\end{minipage} & \begin{minipage}[b]{\linewidth}\raggedleft
2023
\end{minipage} & \begin{minipage}[b]{\linewidth}\raggedleft
2024
\end{minipage} & \begin{minipage}[b]{\linewidth}\raggedleft
2025
\end{minipage} \\
\midrule\noalign{}
\endhead
\bottomrule\noalign{}
\endlastfoot
None (first round) & 3.16 & 4.22 & 5.08 \\
Only ratings on papers that end up accepted & 1.67 & 1.69 & 1.53 \\
Only ratings on papers that end up rejected & 2.99 & 3.88 & 5.15 \\
All (final) & 1.61 & 1.66 & 1.70 \\
\end{longtable}

Letting only the ratings on accepted papers take their final values
brings the noise down to about where the final ratings leave it. Letting
only those on rejected papers do so leaves it near its first-round
level. So the high agreement of final ratings is made on the papers that
end up accepted, by ratings rising there, and final ratings are not a
second, independent reading of the papers. The replay says where the
agreement is made, not whether the rises led to the acceptances or
followed them.

\textbf{If the changes were alignment and nothing else.} What would
happen to the accepted list if each score change depended only on where
the reviews started?

\emph{Method.} We fitted each year's first-round model to the same
decided papers that supply the rating changes: 5,449 in 2024 and 8,541
in 2025, each with at least three paired reviews. The fitted noise is
4.22 and 5.08. In twenty runs of 40,000 papers per year, we draw
first-round ratings with that year's thresholds and distribution of
review counts. For each rating we independently draw its final value
from real reviews with the same initial rating and distance band (the
seven bands above). An empty cell leaves the rating unchanged. We rank
by mean score before and after, selecting the top 32\% each time and
breaking cutoff ties independently. We apply the same rule to the two
observed sets of scores and average over cutoff ties. In the simulation,
a paper belongs if its quality is in the top 32\% of that cohort.

\begin{longtable}[]{@{}
  >{\raggedright\arraybackslash}p{(\linewidth - 8\tabcolsep) * \real{0.3131}}
  >{\raggedleft\arraybackslash}p{(\linewidth - 8\tabcolsep) * \real{0.1717}}
  >{\raggedleft\arraybackslash}p{(\linewidth - 8\tabcolsep) * \real{0.1717}}
  >{\raggedleft\arraybackslash}p{(\linewidth - 8\tabcolsep) * \real{0.1717}}
  >{\raggedleft\arraybackslash}p{(\linewidth - 8\tabcolsep) * \real{0.1717}}@{}}
\caption{Simulated and observed changes among decided papers. Both lists
select the same share by score. The earlier table compares first-round
leaders with area-chair decisions at each year's actual acceptance
rate.}\tabularnewline
\toprule\noalign{}
\begin{minipage}[b]{\linewidth}\raggedright
\end{minipage} & \begin{minipage}[b]{\linewidth}\raggedleft
2024 simulated
\end{minipage} & \begin{minipage}[b]{\linewidth}\raggedleft
2024 observed
\end{minipage} & \begin{minipage}[b]{\linewidth}\raggedleft
2025 simulated
\end{minipage} & \begin{minipage}[b]{\linewidth}\raggedleft
2025 observed
\end{minipage} \\
\midrule\noalign{}
\endfirsthead
\toprule\noalign{}
\begin{minipage}[b]{\linewidth}\raggedright
\end{minipage} & \begin{minipage}[b]{\linewidth}\raggedleft
2024 simulated
\end{minipage} & \begin{minipage}[b]{\linewidth}\raggedleft
2024 observed
\end{minipage} & \begin{minipage}[b]{\linewidth}\raggedleft
2025 simulated
\end{minipage} & \begin{minipage}[b]{\linewidth}\raggedleft
2025 observed
\end{minipage} \\
\midrule\noalign{}
\endhead
\bottomrule\noalign{}
\endlastfoot
Ratings raised / lowered & 21.3\% / 1.4\% & 21.5\% / 1.4\% & 29.7\% /
1.5\% & 29.9\% / 1.5\% \\
Correlation of two reviews, first round to final & 0.17 to 0.27 & 0.17
to 0.33 & 0.15 to 0.26 & 0.15 to 0.33 \\
First-round top 32\% replaced & 19.1\% & 23.3\% & 21.1\% & 26.1\% \\
\end{longtable}

\begin{longtable}[]{@{}
  >{\raggedright\arraybackslash}p{(\linewidth - 4\tabcolsep) * \real{0.6222}}
  >{\raggedleft\arraybackslash}p{(\linewidth - 4\tabcolsep) * \real{0.1889}}
  >{\raggedleft\arraybackslash}p{(\linewidth - 4\tabcolsep) * \real{0.1889}}@{}}
\caption{What alignment alone does to the accepted list
(simulation).}\tabularnewline
\toprule\noalign{}
\begin{minipage}[b]{\linewidth}\raggedright
\end{minipage} & \begin{minipage}[b]{\linewidth}\raggedleft
2024
\end{minipage} & \begin{minipage}[b]{\linewidth}\raggedleft
2025
\end{minipage} \\
\midrule\noalign{}
\endfirsthead
\toprule\noalign{}
\begin{minipage}[b]{\linewidth}\raggedright
\end{minipage} & \begin{minipage}[b]{\linewidth}\raggedleft
2024
\end{minipage} & \begin{minipage}[b]{\linewidth}\raggedleft
2025
\end{minipage} \\
\midrule\noalign{}
\endhead
\bottomrule\noalign{}
\endlastfoot
Accepted papers that belong, before to after & 62.8\% to 59.6\% & 61.3\%
to 57.7\% \\
Accepted papers from the bottom half, before to after & 18.3\% to 21.4\%
& 19.6\% to 23.1\% \\
Papers brought onto the list: share that belong & 32.1\% & 32.5\% \\
Papers pushed off the list: share that belong & 49.2\% & 49.5\% \\
\end{longtable}

\emph{What it shows.} The replay reproduces about three fifths of the
observed gain in agreement and four fifths of the turnover in each year.
The share of selected papers in the top 32\% by quality falls by 3.3 and
3.6 percentage points, with Monte Carlo standard errors of 0.08 and 0.06
points. In both years the papers that entered the list were less likely
to belong than those they displaced.

We also reran the comparison at the shares accepted among the matched
decided papers, 39.81\% in 2024 and 42.99\% in 2025. With that year's
quota applied to both score lists, the share of selected papers in the
corresponding top group by simulated quality fell from 67.6\% to 64.9\%
in 2024 and from 68.0\% to 65.4\% in 2025.

\emph{Limits.} The replay draws each change from first-round scores
alone. Real reviewers can respond to evidence in a rebuttal, and several
reviewers of one paper can respond to the same evidence. We observe
their recorded scores and the eventual decision. We do not observe which
list contains the better papers.

\textbf{The NeurIPS comparison.} The NeurIPS 2021 consistency experiment
put 882 papers through two independent committees, each with rebuttal
and discussion (Beygelzimer et al. 2023). The committees' mean scores
correlated 0.575 on initial scores and 0.586 on final scores, so
cross-committee agreement changed little. That experiment measures
agreement across committees. It gives no independent measure of paper
quality. It is another conference in another year. And the study's
authors note that some authors put much more effort into the responses
for the copy they thought more likely to be accepted, which would tend
to pull the two copies' final scores apart and so work against finding
any gain; we do not know how large the effect was. The same experiment
shows how much a rerun changes the accepted list: about half of it,
which is also the 2014 result, when each committee accepted about half
of what the other accepted (Cortes and Lawrence 2021).

\textbf{How far the step holds.}

\begin{itemize}
\tightlist
\item
  \emph{Ratings change one way, toward the other reviews.} This holds
  for 2024 and 2025, where each review is paired across dates (the table
  of changes by position). For 2023 we pair reviews by their position in
  each paper's list, checked against review lengths where the pairing
  could be recovered (Appendix B). The 2023 counts show the same upward
  lean and the same concentration on accepted papers, but changed
  ratings moved toward the paper's first-round mean only about half the
  time, and we draw nothing from that. The 2023 share applies to the
  matched sample: the papers were matched across dates by the lengths of
  their reviews (4,587 of 4,833 matched), and papers whose reviews were
  edited are harder to match and change rating more often.
\item
  \emph{The rises sit on papers that end up accepted, and the agreement
  of final ratings is made there.} This holds in all three years. It
  says where, not in which order: a raised rating may have helped bring
  the acceptance or followed it, and the data cannot say which.
\item
  \emph{The replay raises agreement and reshuffles the list.} This holds
  in both simulated cohorts, where the score-only changes also reduce
  selection accuracy.
\item
  \emph{Whether the changes carry new information about the papers is
  not shown either way.} A raised rating may be an error corrected or a
  reviewer giving way, and movement in one direction does not settle it:
  a rebuttal only argues for the paper, so a reviewer who honestly
  changes their mind will also mostly raise. What rebuttal and
  discussion do to a paper's quality is not observed and is not claimed
  here; what is observed is ratings and decisions. The NeurIPS
  experiment measures cross-committee agreement. Others find that
  movement depends on whether authors reply: in 2024 and 2025 reviews
  with no author reply were raised 2.7\% of the time, against 27.0\%
  with one (Louis et al. 2026); our counts do not contain that variable.
\end{itemize}

\section{Appendix F. Withdrawals}\label{appendix-f.-withdrawals}

\textbf{Data.} ICLR 2024 and 2025: first-round ratings and each paper's
final status, which gives no date, so some withdrawals may come after
the rebuttal or the decision. ICLR 2023: three dated crawls (5 November
2022, 20 November 2022 and 10 February 2023), which do give dates.

\begin{longtable}[]{@{}
  >{\raggedright\arraybackslash}p{(\linewidth - 8\tabcolsep) * \real{0.3514}}
  >{\raggedleft\arraybackslash}p{(\linewidth - 8\tabcolsep) * \real{0.1622}}
  >{\raggedleft\arraybackslash}p{(\linewidth - 8\tabcolsep) * \real{0.1622}}
  >{\raggedleft\arraybackslash}p{(\linewidth - 8\tabcolsep) * \real{0.1622}}
  >{\raggedleft\arraybackslash}p{(\linewidth - 8\tabcolsep) * \real{0.1622}}@{}}
\caption{Withdrawn and accepted shares by tenth of the first-round mean
score. Papers tied across a decile boundary contribute proportionally to
each side.}\tabularnewline
\toprule\noalign{}
\begin{minipage}[b]{\linewidth}\raggedright
First-round score tenth (low to high)
\end{minipage} & \begin{minipage}[b]{\linewidth}\raggedleft
Withdrawn 2024
\end{minipage} & \begin{minipage}[b]{\linewidth}\raggedleft
Accepted 2024
\end{minipage} & \begin{minipage}[b]{\linewidth}\raggedleft
Withdrawn 2025
\end{minipage} & \begin{minipage}[b]{\linewidth}\raggedleft
Accepted 2025
\end{minipage} \\
\midrule\noalign{}
\endfirsthead
\toprule\noalign{}
\begin{minipage}[b]{\linewidth}\raggedright
First-round score tenth (low to high)
\end{minipage} & \begin{minipage}[b]{\linewidth}\raggedleft
Withdrawn 2024
\end{minipage} & \begin{minipage}[b]{\linewidth}\raggedleft
Accepted 2024
\end{minipage} & \begin{minipage}[b]{\linewidth}\raggedleft
Withdrawn 2025
\end{minipage} & \begin{minipage}[b]{\linewidth}\raggedleft
Accepted 2025
\end{minipage} \\
\midrule\noalign{}
\endhead
\bottomrule\noalign{}
\endlastfoot
0 to 10\% & 49.4\% & 0.4\% & 56.2\% & 0.3\% \\
10 to 20\% & 45.3\% & 0.8\% & 50.9\% & 1.6\% \\
20 to 30\% & 40.7\% & 3.1\% & 44.5\% & 3.8\% \\
30 to 40\% & 32.8\% & 6.9\% & 34.2\% & 6.7\% \\
40 to 50\% & 20.5\% & 13.6\% & 26.0\% & 17.7\% \\
50 to 60\% & 13.8\% & 25.4\% & 20.0\% & 26.4\% \\
60 to 70\% & 7.8\% & 37.4\% & 12.3\% & 36.8\% \\
70 to 80\% & 3.0\% & 58.4\% & 5.4\% & 58.3\% \\
80 to 90\% & 1.4\% & 75.0\% & 2.2\% & 77.3\% \\
90 to 100\% & 0.6\% & 91.5\% & 0.5\% & 92.7\% \\
\end{longtable}

Reviewed papers withdrawn: 20.9\% in 2023, 21.5\% in 2024 and 25.2\% in
2025. Accepted, of all reviewed papers: 31.3\% and 32.2\% in 2024 and
2025; of those not withdrawn, 39.8\% and 43.0\%. The share of withdrawn
papers that come from the bottom half by score is 88\% (2024) and 84\%
(2025).

\textbf{ICLR 2023, by date.} Of the 4,874 papers crawled, 2 were
withdrawn on 5 November 2022, 698 by 20 November 2022 and 1,020 by 10
February 2023 (20.9\%). About two thirds of that year's withdrawals
therefore came within fifteen days of the reviews' release, before any
decision. By tenth of first-round mean score, from low to high, the
share withdrawn within those fifteen days was 31\%, 33\%, 26\%, 22\%,
16\%, 11\%, 3\%, 1\%, 0\% and 0\%, and 14.4\% overall. Early withdrawals
come from the lowest six tenths of the scores, not only from the lowest
tenth.

\textbf{Quality among withdrawn papers.} We fit the ordinal model to
first-round rating patterns, including withdrawals. At each exact mean
score we use the observed withdrawal rate and integrate the fitted
distribution of quality given each rating pattern. Under this score-only
selection rule, 7.8\% of withdrawals in 2024 and 8.9\% in 2025 come from
the top quarter, about 117 and 256 papers. Another way to read the
result is that 6.7\% and 8.9\% of top-quarter papers withdraw. Among all
withdrawn papers, 26.6\% and 28.4\% are in the top half. Authors may
also act on information beyond their scores, so these are conditional
estimates.

\textbf{What withdrawals do to the fitted noise.} On the matched 2023
papers, the first-round noise is 2.61 with withdrawals and 2.92 among
survivors. For 2024 it rises from 3.31 to 4.22, and for 2025 from 3.58
to 5.08. How much does selection by score explain? We weight each
observed rating pattern by the fraction of papers with that exact mean
score that stayed, then refit the same ordinal likelihood. This gives
about 4.2 in 2024 and 4.80 in 2025, close to the observed survivor fits.
Among papers with a lone low review, withdrawal is 22.6\% and 25.9\%,
against 23.2\% and 28.7\% for controls reweighted to the same
quarter-point score bins.

\section{Appendix G. The 2027 trend}\label{appendix-g.-the-2027-trend}

Reviewed papers per year: 972, 1,565, 2,560, 2,978 and 3,375 for ICLR
2018 to 2022 (papers with final ratings), and 4,833, 6,954, 11,423,
18,791 and 40,458 for 2023 to 2027. Growth on the year before was 43\%
(2023), 44\%, 64\%, 65\% and 115\% (2027). We fitted a constant yearly
growth rate to the logarithm of the counts over a chosen run of years
and carried it forward.

\begin{longtable}[]{@{}
  >{\raggedright\arraybackslash}p{(\linewidth - 8\tabcolsep) * \real{0.2239}}
  >{\raggedleft\arraybackslash}p{(\linewidth - 8\tabcolsep) * \real{0.0896}}
  >{\centering\arraybackslash}p{(\linewidth - 8\tabcolsep) * \real{0.1940}}
  >{\centering\arraybackslash}p{(\linewidth - 8\tabcolsep) * \real{0.1940}}
  >{\raggedright\arraybackslash}p{(\linewidth - 8\tabcolsep) * \real{0.2985}}@{}}
\caption{Expected and actual reviewed papers, 2026 and
2027.}\tabularnewline
\toprule\noalign{}
\begin{minipage}[b]{\linewidth}\raggedright
Trend fitted on
\end{minipage} & \begin{minipage}[b]{\linewidth}\raggedleft
Growth a year
\end{minipage} & \begin{minipage}[b]{\linewidth}\centering
2026: expected / actual
\end{minipage} & \begin{minipage}[b]{\linewidth}\centering
2027: expected / actual
\end{minipage} & \begin{minipage}[b]{\linewidth}\raggedright
2027 beyond trend
\end{minipage} \\
\midrule\noalign{}
\endfirsthead
\toprule\noalign{}
\begin{minipage}[b]{\linewidth}\raggedright
Trend fitted on
\end{minipage} & \begin{minipage}[b]{\linewidth}\raggedleft
Growth a year
\end{minipage} & \begin{minipage}[b]{\linewidth}\centering
2026: expected / actual
\end{minipage} & \begin{minipage}[b]{\linewidth}\centering
2027: expected / actual
\end{minipage} & \begin{minipage}[b]{\linewidth}\raggedright
2027 beyond trend
\end{minipage} \\
\midrule\noalign{}
\endhead
\bottomrule\noalign{}
\endlastfoot
2023 to 2025 & 54\% & 17,200 / 18,800 & 26,400 / 40,500 & 14,000 papers,
35\% of the year \\
2023 to 2026 & 58\% & 18,300 / 18,800 & 28,900 / 40,500 & 11,600 papers,
29\% of the year \\
2018 to 2022 & 37\% & 13,600 / 18,800 & 18,700 / 40,500 & 21,800 papers,
54\% of the year \\
\end{longtable}

On the trend of the three or four years before it, 2026 is on trend and
2027 is the outlier. The first-round noise does not jump with the flood:
its rise on the year before is \(+0.58\), \(+0.28\), \(+0.33\) and
\(+0.30\) for 2024 to 2027. What does change in 2027 is the setup
(Appendix C). The size of the excess depends on the years the trend is
fitted on. Counts are reviewed papers and not submissions, and the 2027
count leaves out 492 papers already withdrawn at the snapshot.

\textbf{A reading, not a finding.} Two stages fit this: first a rise in
popularity (2023 to 2025: fast, steady growth under an unchanged scale
and form, with the noise rising and a little more reviewing per paper),
then in 2027 a jump beyond that trend, met with less review per paper.
That better AI writing tools lowered the effort of submitting is a
common explanation for the jump; nothing in these data tests it. ICLR
2023 closed in September 2022, before such tools were widely available.

\textbf{A probe that did not hold.} A two-component fit places its
second component lower in every year, with a larger spread than the main
component in all but 2023 (shares 1\%, 9\%, 23\%, 7\% and 28\% for 2023
to 2027). The 2027 share is close to that year's share beyond trend, but
2025's is already 23\% with no papers beyond trend and 2026's is 7\%, so
the shares are not stable enough to identify the flood with that group.
The noise measured against the whole pool is nearly the same in the
two-group fit (2.67, 3.21, 3.48, 3.78, 4.09).

\textbf{The five-year fit} (the source of Table 1 and Figure 2):

\begin{longtable}[]{@{}
  >{\raggedright\arraybackslash}p{(\linewidth - 12\tabcolsep) * \real{0.0943}}
  >{\raggedleft\arraybackslash}p{(\linewidth - 12\tabcolsep) * \real{0.1132}}
  >{\raggedleft\arraybackslash}p{(\linewidth - 12\tabcolsep) * \real{0.1132}}
  >{\raggedright\arraybackslash}p{(\linewidth - 12\tabcolsep) * \real{0.2075}}
  >{\raggedright\arraybackslash}p{(\linewidth - 12\tabcolsep) * \real{0.2453}}
  >{\raggedleft\arraybackslash}p{(\linewidth - 12\tabcolsep) * \real{0.1132}}
  >{\raggedleft\arraybackslash}p{(\linewidth - 12\tabcolsep) * \real{0.1132}}@{}}
\caption{Five years of first-round noise. Intervals from 300
resamples.}\tabularnewline
\toprule\noalign{}
\begin{minipage}[b]{\linewidth}\raggedright
ICLR
\end{minipage} & \begin{minipage}[b]{\linewidth}\raggedleft
Papers
\end{minipage} & \begin{minipage}[b]{\linewidth}\raggedleft
Reviews per paper
\end{minipage} & \begin{minipage}[b]{\linewidth}\raggedright
Scale
\end{minipage} & \begin{minipage}[b]{\linewidth}\raggedright
First-round noise (95\% interval)
\end{minipage} & \begin{minipage}[b]{\linewidth}\raggedleft
Correlation of two readings
\end{minipage} & \begin{minipage}[b]{\linewidth}\raggedleft
Final-rating noise
\end{minipage} \\
\midrule\noalign{}
\endfirsthead
\toprule\noalign{}
\begin{minipage}[b]{\linewidth}\raggedright
ICLR
\end{minipage} & \begin{minipage}[b]{\linewidth}\raggedleft
Papers
\end{minipage} & \begin{minipage}[b]{\linewidth}\raggedleft
Reviews per paper
\end{minipage} & \begin{minipage}[b]{\linewidth}\raggedright
Scale
\end{minipage} & \begin{minipage}[b]{\linewidth}\raggedright
First-round noise (95\% interval)
\end{minipage} & \begin{minipage}[b]{\linewidth}\raggedleft
Correlation of two readings
\end{minipage} & \begin{minipage}[b]{\linewidth}\raggedleft
Final-rating noise
\end{minipage} \\
\midrule\noalign{}
\endhead
\bottomrule\noalign{}
\endlastfoot
2023 & 4,833 & 3.75 & six levels & 2.73 (2.52 to 2.99) & 0.27 & 1.31 \\
2024 & 6,954 & 3.86 & six levels & 3.31 (3.02 to 3.62) & 0.23 & 1.42 \\
2025 & 11,423 & 4.05 & six levels & 3.58 (3.37 to 3.84) & 0.22 & 1.34 \\
2026 & 18,791 & 3.90 & six levels, different values & 3.91 (3.72 to
4.15) & 0.20 & none \\
2027 & 40,458 & 3.09 & four levels & 4.21 (4.00 to 4.40) & 0.19 &
none \\
\end{longtable}

The rise over the five years, and over 2023 to 2026, is well outside the
intervals. Neighbouring years' intervals overlap, except 2023 and 2024,
which stand 0.03 apart (2.99 against 3.02). Reviewed papers grow 8.4
times over the same years. Final-rating noise for ICLR 2018 to 2025 lies
between 1.1 and 1.7 with no detectable linear trend: 1.35, 1.09, 1.28,
1.66, 1.11, 1.31, 1.44 and 1.34, on the papers with final ratings in the
paper lists. The five-year fit uses a slightly different set of papers
for 2023 to 2025, which is why its final-rating noise for 2024 is 1.42,
not 1.44.

\section{Appendix H. The remedy simulation, and what more reviews
cost}\label{appendix-h.-the-remedy-simulation-and-what-more-reviews-cost}

\textbf{The staged funnel.} In the design of Arbuzov et al. (2026)
(Sections 5 to 7), an automated gate checks form and basic soundness and
passes 80\% of submissions. Each submission supplies one designated
reader. Each paper that passes the gate appears in 20 sorts, at six
minutes per summary. The leading 30\% of all submissions receive three
full reviews apiece. Averaged over all submissions, sorting thus takes
\(0.8 \times 20 \times 6/60 = 1.6\) author hours per paper. Under the
earlier paper's two-hour assumption, full reviews add
\(0.3 \times 3 \times 2 = 1.8\) hours, bringing the funnel to 3.4 hours
per submission, or about 57\% of the six hours for three full reviews
throughout.

At the three hours per full review used here, sorting still takes 1.6
hours per submission and full reviews add 2.7, for 4.3 hours in all.
Across 40,458 papers that is about 174,000 hours, or 87 working years at
2,000 hours a year. The 125,048 reviews actually posted by 6 November
would take about 188 working years on the same assumptions. The earlier
paper prices an optional monitoring audit separately.

\textbf{The simulation.} We reran the earlier paper's simulation with
every peer judgment's noise multiplied by 1.72, the ratio of ICLR 2027's
fitted noise (4.21) to the earlier paper's setting (2.45). A second run
multiplies the automated gate's noise as well. We compare it with
today's process, the mean of three reviews, at the same noise. Our rerun
accepts 25\% of the papers, close to the earlier paper's calibration.
Judgments are continuous, so ICLR's four-level scale does not enter
except where stated.

\begin{longtable}[]{@{}
  >{\raggedright\arraybackslash}p{(\linewidth - 6\tabcolsep) * \real{0.7049}}
  >{\raggedleft\arraybackslash}p{(\linewidth - 6\tabcolsep) * \real{0.0984}}
  >{\raggedleft\arraybackslash}p{(\linewidth - 6\tabcolsep) * \real{0.0984}}
  >{\raggedleft\arraybackslash}p{(\linewidth - 6\tabcolsep) * \real{0.0984}}@{}}
\caption{The staged funnel at ICLR's noise.}\tabularnewline
\toprule\noalign{}
\begin{minipage}[b]{\linewidth}\raggedright
\end{minipage} & \begin{minipage}[b]{\linewidth}\raggedleft
Best quarter accepted
\end{minipage} & \begin{minipage}[b]{\linewidth}\raggedleft
A rerun repeats
\end{minipage} & \begin{minipage}[b]{\linewidth}\raggedleft
Accepted papers from the bottom half
\end{minipage} \\
\midrule\noalign{}
\endfirsthead
\toprule\noalign{}
\begin{minipage}[b]{\linewidth}\raggedright
\end{minipage} & \begin{minipage}[b]{\linewidth}\raggedleft
Best quarter accepted
\end{minipage} & \begin{minipage}[b]{\linewidth}\raggedleft
A rerun repeats
\end{minipage} & \begin{minipage}[b]{\linewidth}\raggedleft
Accepted papers from the bottom half
\end{minipage} \\
\midrule\noalign{}
\endhead
\bottomrule\noalign{}
\endlastfoot
Today's process, earlier paper's noise & 63\% & 51\% & 10.5\% \\
Today's process, ICLR 2027 noise & 57\% & 44\% & 16.0\% \\
Today's process, ICLR 2027 noise, four levels and observed review counts
& 53\% & 40\% & 19.5\% \\
Staged funnel, earlier paper's noise & 81\% & 73\% & 0.8\% \\
Staged funnel, peers at ICLR 2027 noise & 76\% & 67\% & 2.3\% \\
Staged funnel, peers and gate at ICLR 2027 noise & 75\% & 66\% &
2.5\% \\
\end{longtable}

The funnel loses a few points at the higher noise and keeps its lead:
about 18 to 19 points on the first column against today's process at the
same noise without the scale, and more against today's process on ICLR's
scale. Much of the gain comes from many more judgments per paper.

\textbf{Limits.} It is a simulation. Its largest assumption, the earlier
paper's, is that an author sorting short summaries judges about as
noisily as a reviewer reading the full paper; nothing here measures
that. Our rerun covers the configuration that accepts 25\% of papers.
The comparison at a 32\% quota remains untested here. It is one design.
The point it supports is that a process built for this much noise can
sort far better, not that this design is the one to adopt.

\textbf{What more reviews would cost.} Giving every 2027 paper 2026's
number of reviews per paper: \(40{,}458 \times (3.90 - 3.09)\) is about
32,800 more reviews, about 33,000. At three hours a review that is about
98,000 hours, roughly 100,000. At 2,000 working hours a year it is about
49 working years, some fifty, on top of the 188 of the first round as it
was (125,048 reviews at three hours, about 375,000 hours). What it wins
back, at 2027's noise on the four-level scale, is the share of the
accepted list that belongs: from 58.5\% to 60.8\%, 2.3 points (60.9\% in
the second set of runs in Appendix C, 2.5 points; that appendix gives
fewer reviews per paper 2.5 to 3.1 points of the step, depending on
which change is counted first). These are first-round reviews only, at
our working figure of three hours a review; at two hours, the figure of
our earlier paper, the 125,048 reviews are about 250,000 hours, or 125
working years. They leave out reviews written after 6 November 2026,
rebuttal and discussion, area chairs' time, and the time spent writing
the papers.

\section{Appendix I. Limits in
detail}\label{appendix-i.-limits-in-detail}

\textbf{What ``quality'' means.} It means whatever the reviewers of a
paper share, shared taste included, so the figures measure consistency
and not correctness. A judgment every reviewer shares can reproduce well
and still be wrong. The accepted list that ``belongs'' is the list of
papers in the top share by what reviewers share.

\textbf{Where papers land is a model.} No one observes a paper's true
quality. Every figure about where papers land (Table 1, Figures 1 and 6,
Appendix C) is computed from a model of a conference that accepts on
first-round ratings with ties broken at random, fitted to each year's
ratings. ICLR decides after rebuttal and discussion through area chairs;
how far the first round tells ICLR's decisions is in Appendix E. The
ratings, the score changes, the withdrawals and the decisions are
observed. The model's small misfits (Appendix A) carry over. Appendix C
reports the variation between simulated conferences and the Monte Carlo
standard error of their mean.

\textbf{Timing.} The 2026 record uses ratings restored after the
security incident. The 2027 snapshot was taken early in discussion. The
three dated 2023 crawls show noise falling as discussion progressed. We
cannot measure the size or direction of a timing effect for 2027
(Appendix B).

\textbf{Rated withdrawals in the 2027 snapshot.} The main source
excluded 492 papers already marked withdrawn on 6 November. Score Atlas
independently finished collecting public ratings one minute earlier
({songjhPKU} 2026). It recorded 498 withdrawals, 190 with a total of 579
ratings. Within that archive, the 40,452 active rated papers give noise
4.21. Adding the 190 rated withdrawals gives 4.14 across 40,642 papers.
At a 32\% quota, including those withdrawals moves the estimated share
of accepted papers that belong from 58.47\% to 58.68\%. The main series
uses the original snapshot.

\textbf{Why the noise rose.} We do not know. We know that it rose with
submissions, and that reviewers of one paper differ about as much as
before while the distance between papers fell (Appendix D). Whether the
new submissions are mostly middling, or reviewers became reluctant to
give the top ratings, the ratings cannot say.

\textbf{The remedy is a simulation} under our earlier paper's
assumptions (Appendix H). It measures consistency, not correctness, and
rests on an assumption that nothing here measures.

\textbf{Reviews per paper in 2027.} The area-chair guide sought three
completed reviews per paper (ICLR 2026a). The 6 November snapshot
contained 3.09 per rated paper. Later reviews would change the
review-count comparison. At the fitted thresholds, the scale change
costs 1.6 to 2.2 points across the two review-count settings in Appendix
C.

\textbf{The acceptance rate.} The simulations accept 32\% of reviewed
papers, ICLR's rate in 2024 and 2025 (Appendix F). ICLR 2026 accepted
5,355 of 19,525 valid papers, 27.4\% (ICLR 2026 Program Chairs 2026). We
therefore also report a 28\% quota. At 28\% the line sits higher and the
figures of Appendix C change a little, in the same direction and a
little harsher: a top-tenth paper is rejected three times in ten at
2027's setup. The acceptance rate for 2027 is not known yet.

\textbf{Withdrawals.} The 2024 and 2025 archives record final status
without a withdrawal date. Estimated quality among withdrawn papers
depends on the assumption that withdrawal depends on first-round mean
score alone (Appendix F).

\textbf{Rebuttal and discussion.} The data cannot tell a rating raised
by new information from one raised to match the outcome, and they cannot
say which came first, the rise or the decision (Appendix E). What
rebuttal and discussion do to a paper's quality is not observed and not
claimed, and whether the changes carry new information about the papers
has not been measured. The simulation replays changes using scores alone
among decided papers (Appendix E). The 2023 matching across dates is by
review length.

\textbf{What rides on an acceptance.} That an ICLR acceptance is read in
hiring, funding and promotion as a statement about the work is our
statement of the stakes; nothing in our data measures it. The count of
working years rests on our working figure of three hours a review and
2,000 working hours a year, and counts only the reviews already written
by 6 November 2026.

\section{Appendix J. Extended related
work}\label{appendix-j.-extended-related-work}

This paper rests on five choices, and each meets a separate line of
work. It measures agreement in the first round and keeps that apart from
final ratings. It states agreement as one noise figure, fitted from
public counts of ratings, and asks which side of that ratio moved. It
follows the same papers after the first round. It prices the noise, for
a paper and for the accepted list, in a simulated conference. And it
argues for design over patching.

\textbf{The first reading, kept apart.} That reviewers agree weakly is
long known from journals and grants. A meta-analysis of 48 studies of
journal review put mean reliability at .34 (Bornmann et al. 2010), and
an earlier survey across disciplines found reliability low throughout,
with grant referees agreeing more on what lacks merit (Cicchetti 1991).
At ICLR, Tran et al. (2020) read the public scores year by year. They
report Krippendorff's alpha of 56\%, 42\%, 47\% and 39\% for 2017 to
2020, and state as a contribution that consensus among reviewers
decreased over time. Later work gives figures for single years: a
correlation of 0.361 between two reviewers' scores at ICLR 2023
(Zahorodnii et al. 2025), and at NeurIPS 2021 a correlation of 0.575
between two committees' mean initial scores (Beygelzimer et al. 2023).
Two studies carry a series through 2025. Huang (2026) fits a model of
our family to ICLR 2017 to 2025 and estimates the share of score
variance that belongs to the paper: 0.583 in 2017, then between 0.376
and 0.481, with no significant trend. Tang et al. (2026) tabulate the
spread among a paper's reviews for the same years, from a public mirror
of ICLR's scores (González-Márquez and Kobak 2024); it is level from
2022 to 2025 (0.131, 0.136, 0.134 and 0.132 on a unit scale).

So agreement at ICLR has been measured from public data more than once.
It was reported falling from 2017 to 2020, and on final scores the
paper's share of the variance shows no significant trend through 2025;
Huang says plainly that the paper does not claim a growing lottery. At
NeurIPS, Beygelzimer et al. (2023) compared 2021 with 2014 on
whole-process decisions and found no evidence that review had become
noisier as the conference grew.

These series are on final, post-discussion scores, as Huang states, or
they do not say which round their scores come from (Tran et al. 2020;
Zahorodnii et al. 2025). Where the first round has been isolated at
ICLR, it is for 2024 and 2025. Kargaran et al. (2025) archived each
rating when the reviews were released, and report the mean distance
between a paper's reviews before rebuttal: 0.1620 and 0.1623 of the
scale's range in the two years. Louis et al. (2026) work from the same
archive, and Yang et al. (2025) describe the platform that stores such
snapshots.

What changes here is the round. Huang's
\href{https://arxiv.org/html/2610.06591v2}{version 2} distinguishes the
no-discussion counterfactual from what final scores can identify. We
observe first-round ratings, fit their shared variance, and simulate a
specified decision rule on those ratings. The decisions ICLR would
actually have made without discussion remain unobserved. We inherit the
public-data approach of Tran et al.~and Huang, and the archive of
Kargaran et al.~for 2024 and 2025. We add a first-round series in which
the noise rises from 2.73 to 4.21, beside a final-rating series that is
flat. On final ratings the two fits agree. Huang's shares for 2023 to
2025 (0.434, 0.410 and 0.427) correspond to noise of 1.30, 1.44 and 1.34
in our terms; our fits give 1.31, 1.42 and 1.34 (Appendix G). Our
counting rule follows Huang's reasoning too. Huang treats withdrawal as
selection, since papers likely to be rejected withdraw, and shows that
leaving withdrawn papers' scores out lowers the fitted share in every
year after 2017 and gives it a declining trend. We include rated
withdrawals where the source records them (Appendices F and I).

\textbf{One number from public counts, and which side of it moved.} A
model of paper quality plus reviewer noise was put forward in a blog
post on the NeurIPS 2014 experiment (Price 2014), which credits the idea
to Jacob Abernethy. Cortes and Lawrence (2021) fitted one to that year's
ratings, with an offset for each reviewer, and found about half of the
variation in scores subjective. Versions with thresholds, which respect
a rating scale, have been fitted to the ratings of SIGKDD 2009, with
thresholds for each judge (Flach et al. 2009), and to NIH proposals,
with a bias for each reviewer (Johnson 2008). Visscher and Yengo (2022)
write the same generative form for any number of categories and
assessors. Huang (2026) puts this structure on ICLR: an ordered probit
with a paper effect, thresholds for each year, and no reviewer effects.

Two further results bear on reading such a ratio. For grant review,
Erosheva et al. (2021) show that reliability estimated on proposals
restricted in quality will likely be zero under many scenarios, and that
such estimates should not be read as arbitrariness in the review. And at
ICLR 2023 agreement rises with the confidence reviewers state, from a
correlation of 0.24 at confidence 2 to 0.49 at confidence 5 (Zahorodnii
et al. 2025). Mean stated confidence itself was 3.58, 3.64 and 3.67 in
2023 to 2025, which Huang (2026) calls flat; Yang et al. (2025) find
confidence comparatively stable across years, and Jung et al. (2025) put
its average at about 3.60 over ICLR 2017 to 2025, leaving out 2020, for
which confidence is missing.

How reviews are written also changed in these years. Liang, Izzo, et al.
(2024) estimate that between 6.5\% and 16.9\% of review text at four
venues, ICLR 2024 among them, could have been substantially modified by
language models, more so in reviews that state lower confidence. Russo
Latona et al. (2024) estimate that at least 15.8\% of ICLR 2024 reviews
were written with AI assistance, and that in 53.4\% of pairs the
AI-assisted review scores higher than the human one. Sharma et al.
(2026) find that fully generated reviews compress their ratings and do
not discriminate paper quality. Xu et al. (2026) find that one score
carries very different chances of acceptance across research topics. We
list these as changes inside our window. The ratings do not tell us why
the top thinned (Appendix D), and none of these studies tests it.

Most of the earlier fits model reviewers one by one, through offsets,
biases or thresholds for each judge, which calls for knowing who
reviewed what. Huang's does not, and is fitted to final scores. The
range-restriction result concerns subsets of one pool of proposals.

Two things change here. The first is the input. Our fit takes the counts
of a paper's ratings at each level, by maximum likelihood with quality
integrated out (Appendix A), so any year whose ratings are public can be
fitted, 2026 and 2027 included. The second is the decomposition over
time. On the shared scale of 2023 to 2025 the spread among a paper's
reviews goes from 2.27 to 2.18 to 2.05, while the spread between papers
goes from 0.73 to 0.57 to 0.49, and ratings of 8 or 10 fall from 12.4\%
to 8.0\% (Appendix D). This is range restriction happening to a whole
pool over years, where Erosheva et al.~describe it for a selected
subset. We inherit the model and that reading. We add the dated
decomposition, the thinning of high ratings, and the narrowing of the
confidence gap that Zahorodnii et al.~showed for 2023: two confident
reviewers correlated 0.27 against 0.20 for two unconfident ones in 2023,
and 0.20 against 0.16 in 2025 (as noise, 2.7 against 4.0, then 3.9
against 5.1; the gap was no narrower in 2024, so it is one year's
change). Their correlations for 2023 are higher than ours for the same
year (0.361 overall against our 0.24 on recorded first-round ratings);
they do not state the round or the pairing, and we have not reconciled
the two. Two published series support, independently of ours, the
statement that reviewers of one paper differ about as much as before:
Tang et al.'s level spread within papers, and the equal pre-rebuttal
distances of Kargaran et al.~for 2024 and 2025. Yang et al. (2025) also
note, from their plots, that submissions grew most at ratings of 5.6 to
6.6.

\textbf{What happens after the first round.} Earlier studies agree that
scores seldom move and that they move toward the co-reviewers. At NIPS
2016, of 12,154 reviews submitted before the rebuttal, 1,193 changed
score (Shah et al. 2018). At ACL 2018 most reviewers kept their scores,
and the gap between a reviewer's score and the co-reviewers' was the
main explanatory variable of a change (Gao et al. 2019); Huang et al.
(2023) report peer effects among reviewers at ICLR 2022. With
first-round ratings archived for ICLR 2024 and 2025, Kargaran et al.
(2025) report 81\% and 75\% of scores unchanged, 17\% and 23\% raised,
and 1\% lowered in each year. On the same archive Louis et al. (2026)
find 21.0\% of reviewers raising and 1.4\% lowering; among reviews with
an author reply that they study closely, 62.8\% of those below the
paper's mean rise, against 20.2\% of those above it. The convergence
sits in the papers that are accepted: Kargaran et al.'s distance between
a paper's reviews falls by 41\% and 48\% for orals in the two years and
by 7\% and 6\% for rejected papers, and Yang et al. (2025) show the same
pattern by decision tier.

Causes have been tested in part. A randomised trial at ICML 2020 found
no evidence that which reviewer opens the discussion changes acceptance
(Stelmakh et al. 2020), and a randomised study at ICLR 2025 sent
feedback from a language model to more than 20,000 reviews, after which
27\% of the reviewers who received it updated their reviews (Thakkar et
al. 2025). Whether agreement made in discussion carries between
independent panels has been tested as well. At NeurIPS 2021 two
committees' mean scores correlated 0.575 on initial scores and 0.586 on
final ones (Beygelzimer et al. 2023). In grant review, reviewers on one
constructed NIH panel agreed more after discussion while different
panels agreed less (Pier et al. 2017), and for 65 medical grant
proposals the agreement between two panels was the same, a kappa of
0.23, with discussion or with the mean of the reviewers' scores
(Fogelholm et al. 2012). The direction of movement is not the same
everywhere. Evaluators of research projects who were shown others'
scores lowered their own by more after seeing more critical scores than
they raised them after seeing more favourable ones (Lane et al. 2022),
and at ACL 2018 about as many papers had a lowered score (221) as a
raised one (227) (Gao et al. 2019). On withdrawal, Huang (2026) reports
25.3\% of the 2025 cohort withdrawn, about one standard deviation lower
in latent quality. And authors' own view of their papers runs high: at
NeurIPS 2021 their median predicted chance of acceptance was 70\%, when
about 25\% of papers were accepted (Rastogi et al. 2022).

This line takes the review as its unit: which reviews move, and why.
Convergence is measured inside a paper. Kargaran et al.~describe the
rebuttal as a step that changes outcomes for borderline papers. Louis et
al.~tie movement to the author's reply: 2.7\% of reviews without one
were raised, 27.0\% of those with one. They read a review's first-round
position as room to move on a bounded scale, and not as evidence that
the reviewer had decided before the rebuttal.

Our unit is the agreement between reviews across the whole pool, and our
question is what the final figure is made of. The replay answers it:
updating the ratings on accepted papers alone reproduces nearly all of
the gain, updating the low-standing 18\% of reviews alone gives most of
it, and the gain is papers moving apart (the spread between papers goes
from 0.46 to 0.93 in 2024) far more than reviewers closing ranks
(Appendix E). The descriptive facts are Kargaran et al.'s. Our 2024 and
2025 rows are a second analysis of their archive and not independent
confirmation; including withdrawals reproduces their unchanged-score
percentages, while our replay follows decided papers. We add ICLR 2023
from a separate archive, withdrawal by first-round score and, for 2023,
by date, and the replay. The data cannot tell a corrected error from a
reviewer giving way, and Louis et al.'s reading and Lane et al.'s
opposite asymmetry are two reasons to leave that open.

\textbf{What the noise costs.} The cost of noise has been observed
directly a few times. Two committees at NeurIPS, in 2014 and in 2021,
had about half of their accepted papers in common (Cortes and Lawrence
2021; Beygelzimer et al. 2023); Francois (2015) reanalysed the 2014
decisions. Before that, twelve published psychology articles were
resubmitted and eight of the nine that were reviewed were rejected
(Peters and Ceci 1982); 150 NSF proposals were evaluated again by new
reviewers, and funding was found to depend to a significant extent on
chance (Cole et al. 1981); 248 proposals sent to two agencies drew
scores that correlated 0.592 (Hodgson 1997); and at Australia's NHMRC,
59\% of 620 funded grants were sometimes not funded once the variability
of panel scores was taken into account (Graves et al. 2011).

Simulation then took over. From a noise model, Cortes and Lawrence
(2021) expect an accept consistency of about 63\% with three reviewers
when half the variation is subjective. Tran et al. (2020) put the chance
that an accepted ICLR paper is accepted again at 75\% in 2017 and 66\%
in 2020, and Huang (2026) simulates two committees from the fitted
model. Some work scores the panel against a quality that the simulation
knows. Brezis and Birukou (2020) report the share of the truly best
papers that the mean of three referees selects: 60\% of the best 5\%,
70\% to 80\% of the best 10\% and 80\% to 87\% of the best 15\%, across
the two iterations they show. Zhang et al. (2022) and Tang et al. (2026)
give a paper's chance of acceptance as a function of its quality, and
Johnson (2008) the probability that a funded set contains the best
proposals.

The levers have been tried in such models too. More reviewers help a
little: going from two to five raises Tran et al.'s reproducibility by
3\%, going from two to four lowers Huang's disagreement by 5.3 to 7.7
points, Snell (2015) finds five a practical optimum for fellowship
applications, and Zhang et al. (2022) conclude that more than three
reviews are unlikely to repay their burden when authors resubmit. On the
scale, Huang finds that coarsening ICLR 2021's ten-point scores to four
levels lowers the fitted share from 0.376 to 0.345, a difference Huang
cannot tell from zero, and leaves 47.2\% of papers tied at the threshold
when a committee has two reviewers. Visscher and Yengo (2022) find the
loss from scoring in categories very small above five categories, and
outside peer review Preston and Colman (2000) found that scales of two
to four points performed relatively poorly.

These studies take rerun agreement as the outcome, and Cortes and
Lawrence stress that consistency does not imply correctness. Where
quality is known, the size of the noise is set by assumption, or
calibrated on the spread among a paper's reviews at one venue; Huang's
is fitted, to final scores. Zhang et al.~model a game in which authors
resubmit.

Here the noise is the one measured in each year's first round, the
conference modelled is one that decides on those ratings, and the output
is read for a paper: the share of the accepted list that belongs, the
share from the bottom half, the chance at the line, and the chance of a
paper in the top tenth (Appendix C). We then split the 2027 step three
ways. Of 5.4 points, 0.7 come from the noise, 2.5 to 3.1 from fewer
reviews per paper and 1.6 to 2.2 from the scale; the share from fewer
reviews rests on an early snapshot of the 2027 ratings (Appendix I). We
inherit the mechanism from Price and from Cortes and Lawrence, the
accuracy quantities from Brezis and Birukou, Zhang et al.~and Tang et
al., and the experiments on scale and panel size from Huang. Our figure
for more reviews agrees with theirs: a small gain, about two and a half
points, for some fifty working years. It is the kind of diminishing
return that the Spearman--Brown relation gives for added assessors;
Visscher and Yengo use the same relation for the mean of m assessors.
The four levels of ICLR 2027 lie below the range for which Visscher and
Yengo find the loss small. Cortes and Lawrence's caution applies to us
as well: we measure consistency, and in a model of a conference that
accepts on first-round scores.

\textbf{Design over patching.} That growth strains review is the premise
of a line of position papers, each with a remedy (Sculley et al. 2018;
Kim et al. 2025; Chen et al. 2025), while Rogers and Augenstein (2020)
locate part of the problem in a poorly defined reviewing task; Shah
(2022) is an overview of the field. ICLR went from 490 submissions in
2017 to 11,672 in 2025 (Yang et al. 2025). Bergstrom and Gross (2026)
model a loop in which more submissions overtax reviewers, accuracy
drops, and more authors then try their luck. On where the papers come
from, Liang, Zhang, et al. (2024) measure a steady rise in the use of
language models in papers through early 2024, and Contaldi (2026) dates
a break in arXiv's growth to the winter of 2025 to 2026 and reports it
absent from the AI categories themselves. Neither tests what drove ICLR
2027, and neither do we.

Redesigns exist, on paper and in practice. AAAI 2021 ran a two-phase
process that moved reviewing away from papers likely to be rejected and
toward those near the decision boundary (Leyton-Brown et al. 2022). Su
(2021) proposes using authors' own rankings of their papers, Tang et al.
(2026) a voluntary lottery before review, and Zhang et al. (2022)
analyse how many reviews to solicit and where to set the threshold when
authors resubmit. A separate line corrects reviewers' miscalibration
(Tan et al. 2021; Wang and Shah 2018; MacKay et al. 2017); it works on
differences between reviewers, which our model does not represent. The
decisive test also exists. Two committees have taken the same papers
through the whole process at NeurIPS twice, two separate committees were
set up for one track of ESA 2018 (\emph{The {ESA} 2018 Track {B}
Experiment} 2018), and the same NIH applications have been scored by
several constructed panels and by many reviewers (Pier et al. 2017,
2018).

These proposals start from the load on reviewers. We start from a
measured noise, and cost each fix at it before recommending anything:
going back to 2026 or 2023, more reviews, and one staged design, the
funnel of the earlier paper (Arbuzov et al. 2026), rerun at ICLR 2027's
noise. It accepts 76\% of the best quarter, where three reviews at the
same noise accept 57\% (Appendix H). Staged review and the two-committee
test are both inherited. So is a limit of that test, which the authors
of the 2021 experiment state: some authors put more effort into their
responses for the copy they thought likelier to be accepted. The funnel
figure is a simulation under the earlier paper's assumptions.

\textbf{What is and is not new.} Almost every ingredient here has a
precedent. The rating model is of the family Huang fitted to ICLR, with
ancestors in Flach et al.~and Johnson. Reading ICLR's public scores year
by year is Tran et al.'s idea, and they already reported falling
agreement for 2017 to 2020. That most scores never move, that those that
move go up, and that convergence sits in the accepted papers are
Kargaran et al.'s findings for 2024 and 2025, on data we reuse. The
half-repeated accepted list is the NeurIPS experiments'. Simulated
panels have been scored against known quality by Brezis and Birukou,
Zhang et al.~and Tang et al., and Huang has tried coarser scales and
larger committees. Range restriction is Erosheva et al.'s, and the
confidence gap of 2023 Zahorodnii et al.'s. Huang already treats
withdrawal as selection. Staged review has run at AAAI, and the
two-committee test at NeurIPS.

What we have not found elsewhere is the round and the span: the same
model fitted to first-round ratings, the one independent reading, for
ICLR 2023 to 2027, beside a final-rating series in which Huang found no
significant trend and we find none either. On the first-round ratings of
2023 to 2025, the years that share a scale, we show which side of the
ratio moved, as the clear accepts thinned and a paper's reviewers stayed
as far apart, and that the confident reviewers' edge narrowed. By
replay, we show that nearly all of the agreement gained in final ratings
is reproduced by updating scores on the papers that were accepted. The
quantities that say where papers land are inherited; we compute them at
each year's measured first-round noise, and split the 2027 step into
noise, reviews per paper and scale.

We do not show why the noise rose. We measure consistency and not
correctness. The first-round ratings of 2026 and 2027 are early
snapshots, and the cost of fewer reviews in 2027 rests on one. Where
papers land is computed for a conference that decides on its first round
and accepts 32\%, and the remedy is a simulation (Appendix I).


\begin{CSLReferences}{1}{1}
\bibitem[\citeproctext]{ref-arbuzovDSE}
Arbuzov, Mikhail. 2026. {``Decision System Engineering: Managing Bias
and Noise.''} Manuscript submitted for publication.

\bibitem[\citeproctext]{ref-flood}
Arbuzov, Mikhail L., Lee Mosbacker, and Sisong Bei. 2026. \emph{Change
the Funnel, Not Who May Submit: Engineering Peer Review for the Flood}.
\url{https://arxiv.org/abs/2611.01076}.

\bibitem[\citeproctext]{ref-bergstrom2026}
Bergstrom, Carl T., and Kevin Gross. 2026. {``Screening, Sorting, and
the Feedback Cycles That Imperil Peer Review.''} \emph{PLOS Biology} 24
(2): e3003650. \url{https://doi.org/10.1371/journal.pbio.3003650}.

\bibitem[\citeproctext]{ref-beygelzimer2023}
Beygelzimer, Alina, Yann N. Dauphin, Percy Liang, and Jennifer Wortman
Vaughan. 2023. \emph{Has the Machine Learning Review Process Become More
Arbitrary as the Field Has Grown? {T}he {NeurIPS} 2021 Consistency
Experiment}. \url{https://arxiv.org/abs/2306.03262}.

\bibitem[\citeproctext]{ref-bornmann2010}
Bornmann, Lutz, Rüdiger Mutz, and Hans-Dieter Daniel. 2010. {``A
Reliability-Generalization Study of Journal Peer Reviews: A Multilevel
Meta-Analysis of Inter-Rater Reliability and Its Determinants.''}
\emph{PLoS ONE} 5 (12): e14331.
\url{https://doi.org/10.1371/journal.pone.0014331}.

\bibitem[\citeproctext]{ref-brezis2020}
Brezis, Elise S., and Aliaksandr Birukou. 2020. {``Arbitrariness in the
Peer Review Process.''} \emph{Scientometrics} 123 (1): 393--411.
\url{https://doi.org/10.1007/s11192-020-03348-1}.

\bibitem[\citeproctext]{ref-chen2025}
Chen, Nuo, Moming Duan, Andre Huikai Lin, Qian Wang, Jiaying Wu, and
Bingsheng He. 2025. \emph{Position: The Current {AI} Conference Model Is
Unsustainable! {D}iagnosing the Crisis of Centralized {AI} Conference}.
\url{https://arxiv.org/abs/2508.04586}.

\bibitem[\citeproctext]{ref-cicchetti1991}
Cicchetti, Domenic V. 1991. {``The Reliability of Peer Review for
Manuscript and Grant Submissions: A Cross-Disciplinary Investigation.''}
\emph{Behavioral and Brain Sciences} 14 (1): 119--35.
\url{https://doi.org/10.1017/S0140525X00065675}.

\bibitem[\citeproctext]{ref-cole1981}
Cole, Stephen, Jonathan R. Cole, and Gary A. Simon. 1981. {``Chance and
Consensus in Peer Review.''} \emph{Science} 214 (4523): 881--86.
\url{https://doi.org/10.1126/science.7302566}.

\bibitem[\citeproctext]{ref-contaldi2026}
Contaldi, Carlo R. 2026. \emph{Is the {arXiv} Apocalypse Upon Us? {A}
Break in Preprint Growth Coincident with Agentic {AI}}.
\url{https://arxiv.org/abs/2610.34449}.

\bibitem[\citeproctext]{ref-cortes2021}
Cortes, Corinna, and Neil D. Lawrence. 2021. \emph{Inconsistency in
Conference Peer Review: Revisiting the 2014 {NeurIPS} Experiment}.
\url{https://arxiv.org/abs/2109.09774}.

\bibitem[\citeproctext]{ref-erosheva2021}
Erosheva, Elena A., Patrícia Martinková, and Carole J. Lee. 2021.
{``When Zero May Not Be Zero: A Cautionary Note on the Use of
Inter-Rater Reliability in Evaluating Grant Peer Review.''}
\emph{Journal of the Royal Statistical Society Series A: Statistics in
Society} 184 (3): 904--19. \url{https://doi.org/10.1111/rssa.12681}.

\bibitem[\citeproctext]{ref-iclr2023crawl}
{fedebotu}. 2022. \emph{{ICLR} 2023 {OpenReview} Data}.
\url{https://github.com/fedebotu/ICLR2023-OpenReviewData}.

\bibitem[\citeproctext]{ref-flach2009}
Flach, Peter A., Sebastian Spiegler, Bruno Golénia, et al. 2009.
{``Novel Tools to Streamline the Conference Review Process: Experiences
from {SIGKDD}'09.''} \emph{ACM SIGKDD Explorations Newsletter} 11 (2):
63--67. \url{https://doi.org/10.1145/1809400.1809413}.

\bibitem[\citeproctext]{ref-fogelholm2012}
Fogelholm, Mikael, Saara Leppinen, Anssi Auvinen, Jani Raitanen, Anu
Nuutinen, and Kalervo Väänänen. 2012. {``Panel Discussion Does Not
Improve Reliability of Peer Review for Medical Research Grant
Proposals.''} \emph{Journal of Clinical Epidemiology} 65 (1): 47--52.
\url{https://doi.org/10.1016/j.jclinepi.2011.05.001}.

\bibitem[\citeproctext]{ref-francois2015}
Francois, Olivier. 2015. \emph{Arbitrariness of Peer Review: A
{B}ayesian Analysis of the {NIPS} Experiment}.
\url{https://arxiv.org/abs/1507.06411}.

\bibitem[\citeproctext]{ref-gao2019}
Gao, Yang, Steffen Eger, Ilia Kuznetsov, Iryna Gurevych, and Yusuke
Miyao. 2019. {``Does My Rebuttal Matter? {I}nsights from a Major {NLP}
Conference.''} \emph{Proceedings of the 2019 Conference of the North
American Chapter of the Association for Computational Linguistics: Human
Language Technologies, Volume 1 (Long and Short Papers)} (Minneapolis,
Minnesota), 1274--90. \url{https://doi.org/10.18653/v1/N19-1129}.

\bibitem[\citeproctext]{ref-gonzalez2024}
González-Márquez, Rita, and Dmitry Kobak. 2024. \emph{Learning
Representations of Learning Representations}.
\url{https://arxiv.org/abs/2404.08403}.

\bibitem[\citeproctext]{ref-graves2011}
Graves, Nicholas, Adrian G. Barnett, and Philip Clarke. 2011. {``Funding
Grant Proposals for Scientific Research: Retrospective Analysis of
Scores by Members of Grant Review Panel.''} \emph{BMJ} 343: d4797.
\url{https://doi.org/10.1136/bmj.d4797}.

\bibitem[\citeproctext]{ref-hodgson1997}
Hodgson, Corinne. 1997. {``How Reliable Is Peer Review? {A}n Examination
of Operating Grant Proposals Simultaneously Submitted to Two Similar
Peer Review Systems.''} \emph{Journal of Clinical Epidemiology} 50 (11):
1189--95. \url{https://doi.org/10.1016/S0895-4356(97)00167-4}.

\bibitem[\citeproctext]{ref-huang2026}
Huang, Feilian. 2026. \emph{The Review Lottery: Benchmarking an
Observational Estimator of Peer-Review Noise ({ICLR} 2017--2025)}.
Version 2. arXiv, 8 October 2026.
\url{https://arxiv.org/html/2610.06591v2}.
\url{https://doi.org/10.48550/arXiv.2610.06591}.

\bibitem[\citeproctext]{ref-huang2023}
Huang, Junjie, Win-bin Huang, Yi Bu, Qi Cao, Huawei Shen, and Xueqi
Cheng. 2023. {``What Makes a Successful Rebuttal in Computer Science
Conferences?: A Perspective on Social Interaction.''} \emph{Journal of
Informetrics} 17 (3): 101427.
\url{https://doi.org/10.1016/j.joi.2023.101427}.

\bibitem[\citeproctext]{ref-iclr2027ac}
ICLR. 2026a. \emph{{ICLR} 2027 Area Chair Guide}.
\url{https://iclr.cc/Conferences/2027/AreaChairGuidelines}.

\bibitem[\citeproctext]{ref-iclr2027authors}
ICLR. 2026b. \emph{{ICLR} 2027 Author Guidelines}.
\url{https://iclr.cc/Conferences/2027/AuthorGuidelines}.

\bibitem[\citeproctext]{ref-iclr2026incident}
ICLR 2026 Program Chairs. 2025. \emph{{ICLR} 2026 Response to Security
Incident}.
\url{https://blog.iclr.cc/iclr-2026-response-to-security-incident/}.

\bibitem[\citeproctext]{ref-iclr2026retro}
ICLR 2026 Program Chairs. 2026. \emph{A Retrospective on the {ICLR} 2026
Review Process}.
\url{https://blog.iclr.cc/2026/03/31/a-retrospective-on-the-iclr-2026-review-process/}.

\bibitem[\citeproctext]{ref-iclr2027policy}
ICLR 2027 Program Chairs. 2026. \emph{Submission Policies for {ICLR}
2027}.
\url{https://blog.iclr.cc/2026/09/02/submission-policies-for-iclr-2027/}.

\bibitem[\citeproctext]{ref-johnson2008}
Johnson, Valen E. 2008. {``Statistical Analysis of the {N}ational
{I}nstitutes of {H}ealth Peer Review System.''} \emph{Proceedings of the
National Academy of Sciences} 105 (32): 11076--80.
\url{https://doi.org/10.1073/pnas.0804538105}.

\bibitem[\citeproctext]{ref-jung2025}
Jung, Sangkeun, Goun Pyeon, Inbum Heo, and Hyungjin Ahn. 2025.
\emph{What Drives Paper Acceptance? {A} Process-Centric Analysis of
Modern Peer Review}. \url{https://arxiv.org/abs/2509.25701}.

\bibitem[\citeproctext]{ref-kargaran2025}
Kargaran, Amir Hossein, Nafiseh Nikeghbal, Jing Yang, and Nedjma
Ousidhoum. 2025. \emph{Insights from the {ICLR} Peer Review and Rebuttal
Process}. \url{https://arxiv.org/abs/2511.15462}.

\bibitem[\citeproctext]{ref-kim2025}
Kim, Jaeho, Yunseok Lee, and Seulki Lee. 2025. \emph{Position: The {AI}
Conference Peer Review Crisis Demands Author Feedback and Reviewer
Rewards}. \url{https://arxiv.org/abs/2505.04966}.

\bibitem[\citeproctext]{ref-lane2022}
Lane, Jacqueline N., Misha Teplitskiy, Gary Gray, et al. 2022.
{``Conservatism Gets Funded? {A} Field Experiment on the Role of
Negative Information in Novel Project Evaluation.''} \emph{Management
Science} 68 (6): 4478--95. \url{https://doi.org/10.1287/mnsc.2021.4107}.

\bibitem[\citeproctext]{ref-leytonbrown2022}
Leyton-Brown, Kevin, Mausam, Yatin Nandwani, et al. 2022. \emph{Matching
Papers and Reviewers at Large Conferences}.
\url{https://arxiv.org/abs/2202.12273}.

\bibitem[\citeproctext]{ref-liang2024monitoring}
Liang, Weixin, Zachary Izzo, Yaohui Zhang, et al. 2024. \emph{Monitoring
{AI}-Modified Content at Scale: A Case Study on the Impact of {ChatGPT}
on {AI} Conference Peer Reviews}.
\url{https://arxiv.org/abs/2403.07183}.

\bibitem[\citeproctext]{ref-liang2024mapping}
Liang, Weixin, Yaohui Zhang, Zhengxuan Wu, et al. 2024. \emph{Mapping
the Increasing Use of {LLMs} in Scientific Papers}.
\url{https://arxiv.org/abs/2404.01268}.

\bibitem[\citeproctext]{ref-louis2026}
Louis, Mathieu, Tibo Vanleke, Vincent Ginis, and Andres Algaba. 2026.
\emph{Rebuttals Move Peer-Review Scores, but Initial-Review Structure
Bounds the Movement}. \url{https://arxiv.org/abs/2606.22166}.

\bibitem[\citeproctext]{ref-mackay2017}
MacKay, Robert S., Ralph Kenna, Robert J. Low, and Sarah Parker. 2017.
{``Calibration with Confidence: A Principled Method for Panel
Assessment.''} \emph{Royal Society Open Science} 4 (2): 160760.
\url{https://doi.org/10.1098/rsos.160760}.

\bibitem[\citeproctext]{ref-iclrinsightsdata}
Paper Copilot. 2025. \emph{Insights from the {ICLR} Peer Review and
Rebuttal Process (Data Repository)}.
\url{https://github.com/papercopilot/iclr-insights}.

\bibitem[\citeproctext]{ref-paperlists}
Paper Copilot. 2026. \emph{Paper Lists}.
\url{https://github.com/papercopilot/paperlists}.

\bibitem[\citeproctext]{ref-peters1982}
Peters, Douglas P., and Stephen J. Ceci. 1982. {``Peer-Review Practices
of Psychological Journals: The Fate of Published Articles, Submitted
Again.''} \emph{Behavioral and Brain Sciences} 5 (2): 187--95.
\url{https://doi.org/10.1017/S0140525X00011183}.

\bibitem[\citeproctext]{ref-pier2018}
Pier, Elizabeth L., Markus Brauer, Amarette Filut, et al. 2018. {``Low
Agreement Among Reviewers Evaluating the Same {NIH} Grant
Applications.''} \emph{Proceedings of the National Academy of Sciences
of the United States of America} 115 (12): 2952--57.
\url{https://doi.org/10.1073/pnas.1714379115}.

\bibitem[\citeproctext]{ref-pier2017}
Pier, Elizabeth L., Joshua Raclaw, Anna Kaatz, et al. 2017. {``{`Your
Comments Are Meaner Than Your Score'}: Score Calibration Talk Influences
Intra- and Inter-Panel Variability During Scientific Grant Peer
Review.''} \emph{Research Evaluation} 26 (1): 1--14.
\url{https://doi.org/10.1093/reseval/rvw025}.

\bibitem[\citeproctext]{ref-preston2000}
Preston, Carolyn C., and Andrew M. Colman. 2000. {``Optimal Number of
Response Categories in Rating Scales: Reliability, Validity,
Discriminating Power, and Respondent Preferences.''} \emph{Acta
Psychologica} 104 (1): 1--15.
\url{https://doi.org/10.1016/S0001-6918(99)00050-5}.

\bibitem[\citeproctext]{ref-price2014}
Price, Eric. 2014. \emph{The {NIPS} Experiment}. Guest post on the blog
Moody Rd.
\url{https://blog.mrtz.org/2014/12/15/the-nips-experiment.html}.

\bibitem[\citeproctext]{ref-rastogi2022}
Rastogi, Charvi, Ivan Stelmakh, Alina Beygelzimer, et al. 2022.
\emph{How Do Authors' Perceptions of Their Papers Compare with
Co-Authors' Perceptions and Peer-Review Decisions?}
\url{https://arxiv.org/abs/2211.12966}.

\bibitem[\citeproctext]{ref-rogers2020}
Rogers, Anna, and Isabelle Augenstein. 2020. {``What Can We Do to
Improve Peer Review in {NLP}?''} \emph{Findings of the Association for
Computational Linguistics: EMNLP 2020}, 1256--62.
\url{https://doi.org/10.18653/v1/2020.findings-emnlp.112}.

\bibitem[\citeproctext]{ref-latona2024}
Russo Latona, Giuseppe, Manoel Horta Ribeiro, Tim R. Davidson, Veniamin
Veselovsky, and Robert West. 2024. \emph{The {AI} Review Lottery:
Widespread {AI}-Assisted Peer Reviews Boost Paper Scores and Acceptance
Rates}. \url{https://arxiv.org/abs/2405.02150}.

\bibitem[\citeproctext]{ref-sculley2018}
Sculley, D., Jasper Snoek, and Alex Wiltschko. 2018. \emph{Avoiding a
Tragedy of the Commons in the Peer Review Process}.
\url{https://arxiv.org/abs/1901.06246}.

\bibitem[\citeproctext]{ref-shah2022}
Shah, Nihar B. 2022. {``Challenges, Experiments, and Computational
Solutions in Peer Review.''} \emph{Communications of the ACM} 65 (6):
76--87. \url{https://doi.org/10.1145/3528086}.

\bibitem[\citeproctext]{ref-shah2018}
Shah, Nihar B., Behzad Tabibian, Krikamol Muandet, Isabelle Guyon, and
Ulrike von Luxburg. 2018. {``Design and Analysis of the {NIPS} 2016
Review Process.''} \emph{Journal of Machine Learning Research} 19 (49):
1--34. \url{https://jmlr.org/papers/v19/17-511.html}.

\bibitem[\citeproctext]{ref-sharma2026}
Sharma, Vibhhu, Thorsten Joachims, and Sarah Dean. 2026. \emph{Do {LLMs}
Favor {LLMs}? {Q}uantifying Interaction Effects in Peer Review}.
\url{https://arxiv.org/abs/2601.20920}.

\bibitem[\citeproctext]{ref-iclrscores}
{shiqiangw}. 2026. \emph{{ICLR} 2027 Scores}.
\url{https://shiqiangw.github.io/iclr-2027-scores/}.

\bibitem[\citeproctext]{ref-snell2015}
Snell, Richard R. 2015. {``Menage a Quoi? {O}ptimal Number of Peer
Reviewers.''} \emph{PLOS ONE} 10 (4): e0120838.
\url{https://doi.org/10.1371/journal.pone.0120838}.

\bibitem[\citeproctext]{ref-scoreatlas}
{songjhPKU}. 2026. \emph{{ICLR} 2027 Score Atlas: Public Review
Statistics}.
\url{https://songjhpku.github.io/ICLR2027-statistics/en.html}.

\bibitem[\citeproctext]{ref-stelmakh2020}
Stelmakh, Ivan, Charvi Rastogi, Nihar B. Shah, Aarti Singh, and Hal
Daumé III. 2020. \emph{A Large Scale Randomized Controlled Trial on
Herding in Peer-Review Discussions}.
\url{https://arxiv.org/abs/2011.15083}.

\bibitem[\citeproctext]{ref-su2021}
Su, Weijie J. 2021. \emph{You Are the Best Reviewer of Your Own Papers:
An Owner-Assisted Scoring Mechanism}.
\url{https://arxiv.org/abs/2110.14802}.

\bibitem[\citeproctext]{ref-tan2021}
Tan, Sijun, Jibang Wu, Xiaohui Bei, and Haifeng Xu. 2021. {``Least
Square Calibration for Peer Reviews.''} \emph{Advances in Neural
Information Processing Systems} 34.
\url{https://papers.nips.cc/paper/2021/hash/e354fd90b2d5c777bfec87a352a18976-Abstract.html}.

\bibitem[\citeproctext]{ref-tang2026}
Tang, Theo, Toby Handfield, and Julian Garcia. 2026. \emph{Can We
Volunteer Out of the Peer Review Crisis?}
\url{https://arxiv.org/abs/2604.27900}.

\bibitem[\citeproctext]{ref-thakkar2025}
Thakkar, Nitya, Mert Yuksekgonul, Jake Silberg, et al. 2025. \emph{Can
{LLM} Feedback Enhance Review Quality? {A} Randomized Study of 20{K}
Reviews at {ICLR} 2025}. \url{https://arxiv.org/abs/2504.09737}.

\bibitem[\citeproctext]{ref-bast2018}
\emph{The {ESA} 2018 Track {B} Experiment}. 2018. Project page, Chair of
Algorithms and Data Structures, University of Freiburg.
\url{https://ad-wiki.informatik.uni-freiburg.de/research/ESA2018Experiment}.

\bibitem[\citeproctext]{ref-tran2020}
Tran, David, Alex Valtchanov, Keshav Ganapathy, et al. 2020. \emph{An
Open Review of {OpenReview}: A Critical Analysis of the Machine Learning
Conference Review Process}. \url{https://arxiv.org/abs/2010.05137}.

\bibitem[\citeproctext]{ref-visscher2022}
Visscher, Peter M., and Loic Yengo. 2022. {``The Effect of the Scale of
Grant Scoring on Ranking Accuracy.''} \emph{F1000Research} 11: 1197.
\url{https://doi.org/10.12688/f1000research.125400.1}.

\bibitem[\citeproctext]{ref-wang2018}
Wang, Jingyan, and Nihar B. Shah. 2018. \emph{Your 2 Is My 1, Your 3 Is
My 9: Handling Arbitrary Miscalibrations in Ratings}.
\url{https://arxiv.org/abs/1806.05085}.

\bibitem[\citeproctext]{ref-xu2026}
Xu, Binyan, Xilin Dai, Fan Yang, and Kehuan Zhang. 2026. \emph{Reviewer
Scores Are Not Comparable Across Research Areas in {ML} Peer Review}.
\url{https://arxiv.org/abs/2607.27209}.

\bibitem[\citeproctext]{ref-yang2025}
Yang, Jing, Qiyao Wei, and Jiaxin Pei. 2025. \emph{Paper Copilot:
Tracking the Evolution of Peer Review in {AI} Conferences}.
\url{https://arxiv.org/abs/2510.13201}.

\bibitem[\citeproctext]{ref-zahorodnii2025}
Zahorodnii, Andrii, Jasper J. F. van den Bosch, Ian Charest, Christopher
Summerfield, and Ila R. Fiete. 2025. \emph{Paper Quality Assessment
Based on Individual Wisdom Metrics from Open Peer Review}.
\url{https://arxiv.org/abs/2501.13014}.

\bibitem[\citeproctext]{ref-zhang2023system}
Zhang, Yichi, Fang-Yi Yu, Grant Schoenebeck, and David Kempe. 2022. {``A
System-Level Analysis of Conference Peer Review.''} \emph{Proceedings of
the 23rd {ACM} Conference on Economics and Computation ({EC} '22)},
1041--80. \url{https://doi.org/10.1145/3490486.3538235}.

\end{CSLReferences}
\end{document}